\documentclass[aps,showpacs,twocolumn,superscriptaddress]{revtex4-2}
\usepackage{graphicx}% Include figure files
\usepackage{dcolumn}% Align table columns on decimal point
\usepackage{bm}% bold math
\usepackage{amsmath,amsthm,amsfonts,amssymb,amscd,epsfig, enumerate,yfonts}
\usepackage{tikz}
\usepackage{ragged2e}
\usepackage{booktabs}
\usepackage{mathtools}
\usepackage{multirow}
\usepackage{color}
\usepackage[normalem]{ulem} % \sout{old text} for strikeout
\usepackage[dvipsnames]{xcolor} % For blue in-text comments
\usepackage{natbib}
\usepackage{hyperref}
\hypersetup{
  colorlinks=true,        % false: boxed links; true: colored links
  linkcolor=blue,         % color of internal links
  citecolor=cyan,         % color of links to bibliography
}

\begin{document}

\title{Probing the nature of Einstein–nonlinear Maxwell–Yukawa black hole through gravitational wave forms from periodic orbits and quasiperiodic oscillations }

\author{Oreeda Shabbir}
\email{oreedashabbir7@gmail.com}
\affiliation{School of Natural Sciences, National University of Sciences and Technology (NUST), Islamabad, 44000, Pakistan}

\author{Abubakir Shermatov % \orcidicon{0009-0009-4044-4507}
}
\email{shermatov.abubakir98@gmail.com}
\affiliation{%Department of Information Technologies, 
Tashkent International University of Education, Imom Bukhoriy 6, Tashkent 100207, Uzbekistan}
\affiliation{University of Tashkent for Applied Sciences, Gavhar Str. 1, Tashkent 700127, Uzbekistan}
\affiliation{Institute of Fundamental and Applied Research, National Research University TIIAME, Kori Niyoziy 39, Tashkent 100000, Uzbekistan}
\affiliation{Tashkent State Technical University, Tashkent 100095, Uzbekistan}

\author{Bushra Majeed}
\email{bushra.majeed@ceme.nust.edu.pk}
\affiliation{College of Electrical and Mechanical Engineering (CEME), National University of Sciences and Technology, Islamabad, 44000, Pakistan}
\author{Tehreem Zahra}
\email{tehreemzahra971@gmail.com}
\affiliation{Center for Astronomy and Astrophysics, Department of Physics, Fudan University, Shanghai 200438, China}
\author{Mubasher Jamil}
\email{mjamil@candqrc.ca}
\affiliation{School of Natural Sciences, National University of Sciences and Technology (NUST), Islamabad, 44000, Pakistan}
\affiliation{Research Center of Astrophysics and Cosmology, Khazar University, Baku, AZ 1096, 41 Mehseti Street, Azerbaijan}

\author{Javlon Rayimbaev %\orcidicon{0000-0001-9293-1838}
}
\email{javlon@astrin.uz}
%\affiliation{Institute of Fundamental and Applied Research, National Research University TIIAME, Kori Niyoziy 39, Tashkent 100000, Uzbekistan}
\affiliation{National University of Uzbekistan, Tashkent 100174, Uzbekistan}
\affiliation{Urgench State University, Kh. Alimdjan str. 14, Urgench 220100, Uzbekistan} 

\date{\today}

\begin{abstract}
In this work, we study gravitational-wave (GW) emission from periodic orbits of test particles, analyze quasi-periodic oscillations (QPOs), and constrain the parameters of the static, spherically symmetric Einstein–nonlinear Maxwell–Yukawa (ENLMY) black hole (BH). Using the Hamiltonian approach, we calculate the equations of motion of the particles. We analyze the effective potential to determine the innermost stable circular orbit (ISCO) and innermost bound circular orbit (IBCO), illustrating how the Yukawa screening parameter $\alpha$ and electric charge $Q$ affect orbital stability and energy requirements. Periodic orbits are classified by integer triplets and exhibit characteristic zoom–whirl behavior. Based on these orbits, we compute the corresponding GW signals in both the $h_+$ and $h_{\times}$ polarizations. Finally, we perform Monte Carlo Markov Chain (MCMC) simulations to constrain the parameters of the ENLMY BH for four microquasars and the galactic center within the relativistic precession model.
\vspace{6pt}

\end{abstract}

\maketitle

\section{Introduction}

Black holes (BHs) are one of the most fascinating and mysterious
objects in the universe, which have been investigated for their unique properties and entangled structures with their environment. The compelling contribution to resolving the
mysteries associated with the BHs took place in 2019. A tremendous
achievement of the Event Horizon Telescope (EHT) is that it obtained the
 first-ever image of the shadow of the event horizon as well as accretion flow encircling a supermassive
BH in the center of M87* galaxy \cite{2019ApJ...875L...1E}. This discovery has enabled the researchers to explore 
the studies of BHs by comparing the existing theoretical models concerning accretion processes near BHs
with the observational data. 
Quasiperiodic oscillations (QPOs) are a powerful tool for testing gravity theories. This study is insightful for the properties of gravitational and
electromagnetic field surrounding BHs, and obtaining the accurate values for the
BH metric parameters.

The Reissner--Nordström (RN) solution, representing a static, spherically symmetric and electrically charged BH in Einstein-Maxwell theory, has long served as a fundamental benchmark in general relativity. Its well-understood geodesic structure provides a clear picture of how electric charge influences the spacetime around a compact object. Significant attention has shifted towards theories of nonlinear electrodynamics, which generalize the Maxwell action and can lead to novel gravitational phenomena \cite{bonn}. A primary motivation for such NED couplings is to explore scenarios in which the classical singularities of point charges or BHs might be resolved, or in which new observational signatures could emerge, as they introduce additional physical parameters that can significantly alter the spacetime structure, leading to phenomena absent in their RN counterparts.
\par
The Yukawa potential, $\phi(r) = \frac{q}{r} e^{-\alpha r}$, generalizes the Coulomb potential through an exponential damping term, where $q$ is the electric charge and $\alpha$ is a positive constant \cite{193548}. This structure introduces a screening mechanism that confines electromagnetic interactions to a finite range, a feature that arises from its description of nuclear forces mediated by massive mesons. The parameter $\alpha$ directly determines this range; larger values lead to faster decay. This makes the potential particularly relevant for modeling screened electromagnetic interactions in contexts like massive gravity and dense astrophysical environments. Crucially, it reduces to the standard Coulomb potential in the limit $\alpha \rightarrow 0$, ensuring consistency with established theory.
\par
Motivated by these theoretical frameworks, we investigate how the modified spacetime geometry of a charged BH, under the influence of a Yukawa-type screened electromagnetic potential, behaves in the strong-field region. While our recent work \cite{2025EPJC...85.1340Z} explored the orbital dynamics and observational signatures of such an $f(R)$-based Yukawa BH \cite{yukawaMetric,yukawa2,yukawa3,yukawa4}, the present study investigates a Yukawa-modified RN metric, which emerges from a nonlinear electrodynamic coupling \cite{Mazharimousavi:2019ksw}. Quasinormal modes (QNMs) and shadows of rotating BHs with a Yukawa-type scalar field in scalar-tensor-vector gravity are studied in Ref.\cite{2025PDU....5002124Z}, providing tests using data from EHT observations.

A key observational signature for probing these modifications arises from the gravitational radiation emitted by compact objects in extreme-mass-ratio inspirals (EMRIs), where a stellar-mass object moves on a bound, slowly decaying orbit around a central massive black hole. In particular, periodic orbits, which often exhibit distinctive zoom-whirl behavior, play an important role in studying the long-term stability of bound systems and provide valuable insights into the evolution of EMRIs, as they serve as transitional paths during the adiabatic inspiral phase. In this Yukawa-modified spacetime, the parameters $\alpha$ and the charge $Q$ are expected to alter the geodesic structure, thereby modifying the characteristics of bound orbital motion. We therefore map the timelike periodic geodesic structure in this Yukawa-modified spacetime and identify the parameter space of energy and angular momentum that supports stable bound orbits.
\par
QPOs are observed as narrow peaks in the power-density spectra of X-ray binaries. The exact origin of QPOs remains unknown, and various models have been proposed to explain them \cite{model1,model2,model3}. In the relativistic precession (RP) model \cite{Stella:Proposed-RPM}, for instance, the observed frequencies are identified with the orbital, periastron precession, and nodal precession frequencies of test particles on nearly circular orbits \cite{Shaymatov_2022, Rayimbaev_2022, Ghasemi_Nodehi_2020, Jusufi_2021, Azreg_A_nou_2020}. The introduction of a Yukawa-type potential modifies these characteristic frequencies, shifting the QPO relations in a parameter-dependent manner. By applying the RP model to well-measured QPO triplets from microquasars, we can therefore obtain simultaneous constraints on the BH mass, spin, charge $Q$, and $\alpha$. Markov Chain Monte Carlo (MCMC) \cite{emcee} analyses of such observational data offer a powerful statistical tool to derive posterior bounds on these parameters. 

We have applied this method to study the gravitational waves from the periodic orbits around a charged black hole with a Yukawa-form scalar field, QPO tests, and to get the constraints on the parameters of the BH in the
relativistic precession model and QPO orbits. For the detailed analysis of QPOs, Monte Carlo
Markov Chain (MCMC) simulation observed at
the center of galaxies M82 and the Milky Way, in the microquasars
XTE J1550-564, GRO J1655-40, and GRS 1915-105 are used.

A rich literature is present where constraints on the parameters of BH are found using MCMC by utilizing the QPOs data of XRBs within the RP model \cite{Motta:precise-mass,Motta:XTE226,Motta:XTE564,Liu-etal2023,2023arXiv231112423L}.
\par
The paper is organized as follows: Section II briefly reviews the ENLMY metric. Section III is dedicated to deriving the equations of motion and the effective potential for particle orbits. In Section IV, we present a detailed analysis of periodic orbits, and in Section V, we analyze the gravitational radiation emitted by periodic orbits of a test particle. In Section VI, we explore the fundamental frequencies and their implications for QPOs, which are discussed in Section VII. We also perform MCMC analysis in Section VIII to find the constraints of four XRBs. Finally, we will provide a brief conclusion to our results. Throughout the paper, we use the metric signature $(-,+,+,+)$ and units in which $G = c = 1$, unless stated. 

\section{The Einstein--Nonlinear Maxwell--Yukawa Metric}

To model deviations from the standard Maxwell electrodynamics, one may couple gravity to a nonlinear electrodynamic Lagrangian constructed from the Yukawa-type potential. Starting from the potential
\begin{equation}
\phi(r) = \frac{q}{r} e^{-\alpha r},
\end{equation}
and following the ENLMY formalism~\cite{Mazharimousavi:2019ksw}, one arrives at the static and spherically symmetric solution given as
\begin{equation}\label{eq.ds}
ds^2 = - f(r)\, dt^2 + \frac{dr^2}{f(r)} + r^2 \left(d\theta^2 + \sin^2\theta\, d\varphi^2\right),
\end{equation}
with the metric function given by

\begin{eqnarray}\label{ENLMYmetric}
    f(r) &=& 1 - \frac{2M}{r} - \frac{Q^2}{6r^2}B(r),\\
\nonumber B(r)&=&\alpha^3 r^3 - \big(\alpha^2 r^2 - 2\alpha r + 6\big)e^{-\alpha r} + \alpha^4 r^4 \mathcal{E}_1(\alpha r),
\end{eqnarray}
where $M$ is the ADM mass, $Q$ is the electric charge, $\alpha > 0$ denotes the Yukawa screening parameter, and $\mathcal{E}_1(x)$ is the exponential integral function defined as~\cite{Mazharimousavi:2019ksw}
\begin{equation}
\mathcal{E}_{1}(x) = \int_{1}^{\infty} \frac{e^{-xt}}{t}\, dt .
\end{equation}

In the limit $\alpha \to 0$, the above solution \eqref{ENLMYmetric}, reduces to the familiar RN BH, confirming its consistency with standard Maxwell electrodynamics. The Yukawa screening introduces short-distance modifications, and depending on the values of $(M,Q,\alpha)$, the spacetime can admit two horizons, an extremal configuration, or describe a naked singularity.

\begin{figure}[ht]
    \centering
    \includegraphics[width=1.0\linewidth]{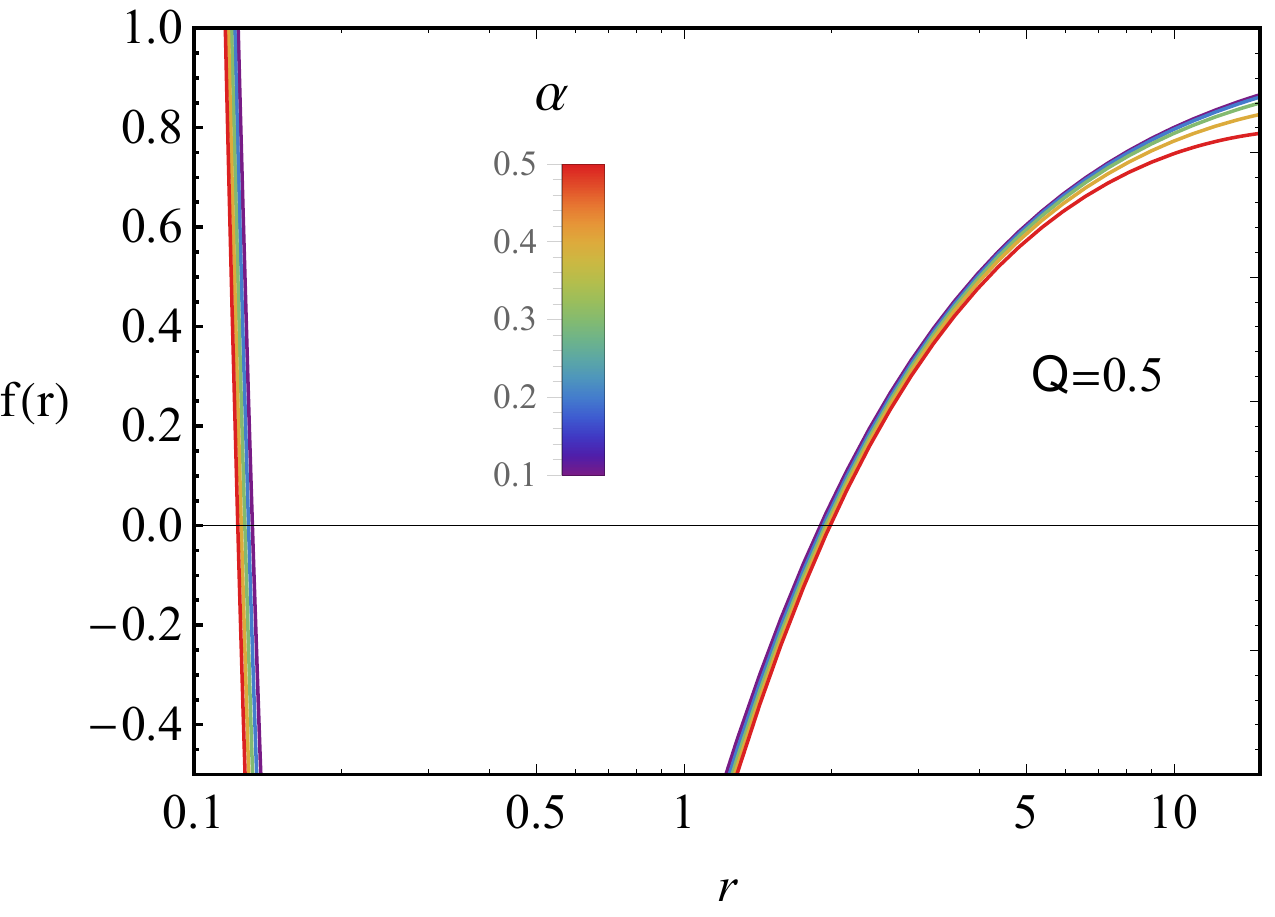}
    \caption{Plot of the metric function $f(r)$ for fixed charge $Q=0.5$ and mass $M=1$ with various values of the Yukawa screening parameter $\alpha$.
    }
    \label{f1}
\end{figure}
\begin{figure}[ht]
    \centering
    \includegraphics[width=1.0\linewidth]{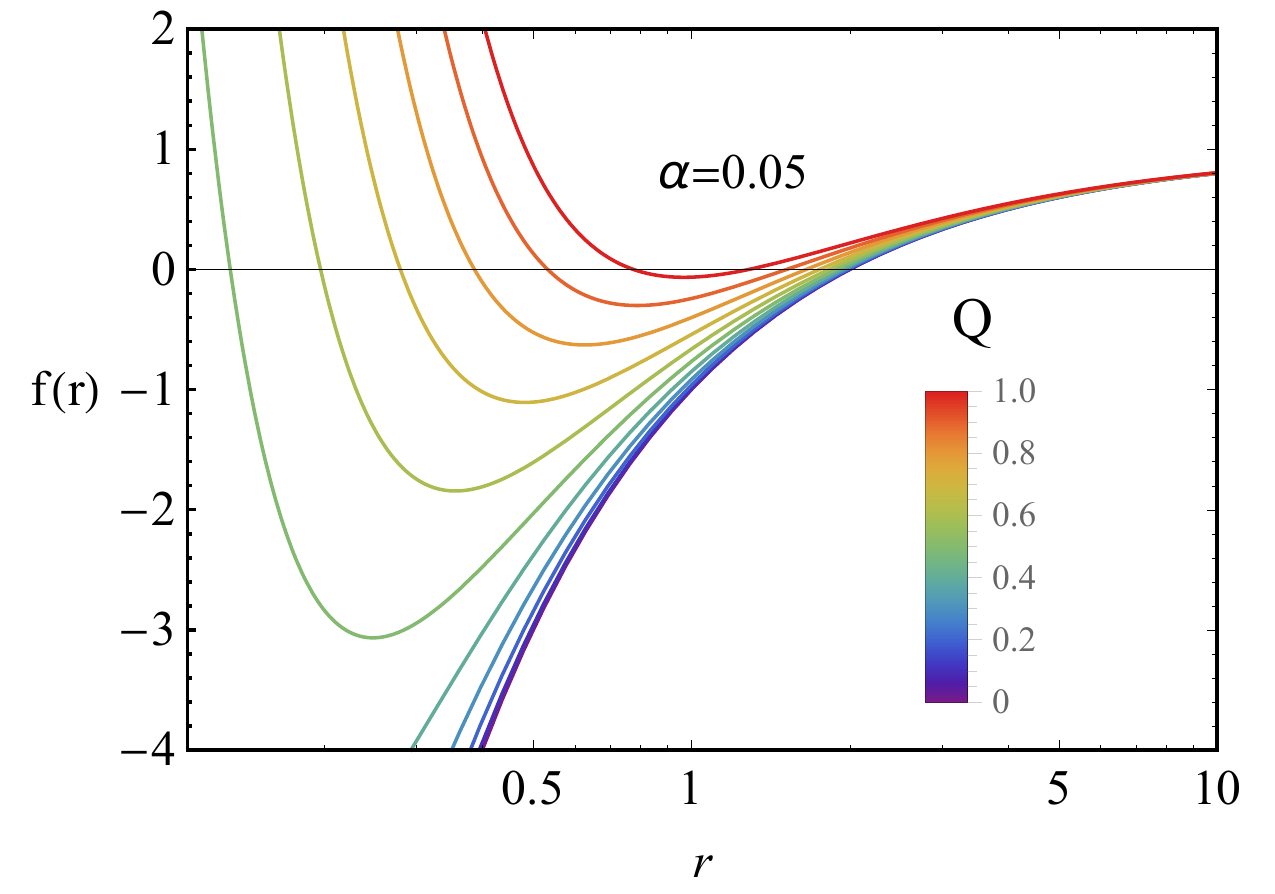}
    \caption{Plot of the metric function $f(r)$ for fixed Yukawa screening parameter $\alpha=0.05$ and mass $M=1$ with various values of $Q$.
    }
    \label{f2}
\end{figure}

In Fig.~\ref{f1}, the curves show that increasing the screening parameter $\alpha$ weakens the effective gravitational potential, causing the metric function $f(r)$ to rise more slowly near the BH and shifting the event horizon to larger radii. This behavior indicates that Yukawa-type screening weakens gravity at short distances. In Fig.~\ref{f2}, we fix $\alpha=0.05$ and vary the charge $Q$. As $Q$ increases, the repulsive contribution associated with the charge becomes more significant, leading to a shallower gravitational well and a reduction in the horizon radius. For sufficiently large values of $Q$, the metric approaches the extremal limit, where the outer and inner horizons tend to merge. Together, these figures demonstrate how the parameters $\alpha$ and $Q$ influence the geometric structure of spacetime and shift the locations of the BH horizons.

\section{Bound orbits of the ENLMY metric}
In this section, we examine the bound orbital structure of ENLMY BH. We focus on the motion of neutral test particles and, in the subsections that follow, we derive the geodesic equations, construct the corresponding effective potential, and determine the radii of the innermost stable circular orbit (ISCO) and the innermost bound circular orbit (IBCO).

\subsection{Geodesic equations}
A neutral test particle moving in the vicinity of a BH can be described using the Hamiltonian formalism \cite{2002clme.book.....G}, given by 
\begin{align}
\mathcal 
H=\frac{1}{2} g^{\zeta \eta} p_{\zeta} p_{\eta} + \frac{1}{2}\mu^2,\label{Ham}
\end{align}
where $\mu$, $p^{\zeta}=\mu 
u^{\zeta}$ represents the rest mass and the four-momentum. The four-velocity $u^{\zeta}= d x^{\zeta}/d\tau$ of the test particle is defined with respect to the proper time $ \tau $ along the particle’s worldline. Hamilton's equations govern the motion of the particle \cite{2002clme.book.....G}
\begin{align}\label{ham.eq}
\frac{dx^{\eta}}{d\zeta}\equiv \mu u^{\eta}=\frac{\partial \mathcal H}{\partial p_{\eta}}, \quad
\frac{d p_{\eta}}{d\zeta} = -\frac{\partial \mathcal H}{\partial x^{\eta}},
\end{align}
with  $ \zeta=\tau/\mu$ as the affine parameter. For massive particles, $\mu \neq 0$ (typically $\mu=1$ in natural units), while for massless particles (e.g., photons), one takes the limit $\mu\rightarrow0$, in which case the motion is described using a null geodesic. Since the BH geometry is symmetric, there are two conserved quantities linked with the symmetries of the metric, named as the specific energy $\mathcal{E}=E/m$, and specific angular momentum,  $\mathcal{L}=L/m$ given by
\begin{align}\label{EE}
    \frac{p_{t}}{\mu}&= -f(r)\frac{dt}{d\tau} =-\mathcal{E}, \\\label{LL}
\frac{p_{\phi}}{\mu}&=
  r^2 \sin^2\theta \, \frac{d \phi}{d \tau} =\mathcal{L}.
\end{align}
  The four-velocity $u^\alpha$ of the particle, written in component form, is given by  
\begin{align}
\frac{dt}{d\tau} &= \frac{\mathcal{E}}{f(r)},\qquad
\frac{d\phi}{d\tau} = \frac{\mathcal{L}}{r^2 \sin^2\theta},\\
\left(\frac{dr}{d\tau}\right)^2&=\mathcal{E}^2- \left(\epsilon + \frac{L^2}{r^2 \sin^2\theta} \right) {f(r)},\nonumber\\
\end{align}
named as the time component $u^t$, azimuthal component, $u^\phi$ and radial component $u^r$. 
The parameter $\epsilon$ takes the value $\epsilon=1$ for time-like particles and $\epsilon=0$ for null-like particles. A dot represents differentiation with respect to the proper time $\tau$.

The Hamiltonian Eq.\eqref{Ham}, for the ENLMY BH~\eqref{eq.ds} can be written in the form
\begin{align}\label{Ham.}
\mathcal H = \frac{1}{2} {f(r)} p_r^2  +\frac{1}{2r^2} p_\theta^2 + \frac{1}{2f(r)} \left[ V_{\rm eff}(r,\theta) - \mathcal{E}^2 \right],
\end{align}
where the effective potential $V_{\text{eff}}(r, \theta)$ takes the form
\begin{align}
\label{Veff}
V_{\text{eff}}(r, \theta) = \left(\mu^2 + \frac{\mathcal{L}^2 \csc^2 \theta }{r^2} \right) {f(r)}.
\end{align}
Using Hamilton’s equations~\eqref{ham.eq}, the equations governing the particle’s motion are given by
\begin{align}
    \dot{p}_t&=\dot{p}_\phi=0,\\
    \dot{p}_\theta &=\frac{1}{2r^2}\partial_\theta\left[\mathbf{\Theta}(\theta)\right],\\
     \dot{p}_r&=\frac{1}{2}\left[-p_r^2 \partial_r\left(f(r)\right)+\frac{2p_\theta^2}{r^3}+\partial_r\left(\frac{\mathcal{R}(r)}{f(r)}\right)-\frac{2\mathbf{\Theta}(\theta)}{r^3}\right],
     \end{align}
     where 
\begin{align}\label{constants}
    \mathcal{R}(r) &={\cal E}^2-\left[f(r)\left(\mu^2+\frac{\mathcal{L}^2}{r^2}+\frac{\mathcal{K}}{r^2}\right)\right],\\
    \mathbf{\Theta}(\theta) &=\mathcal{K}-{{\mathcal L}}^2\cot^2\theta.
    \end{align}
Here, $\mathcal{K}$ is the separation constant. From this point onward, we consider the motion of massive test particles $(\mu=1)$ confined to the equatorial plane by setting $\theta=\frac{\pi}{2}$ and choosing $\mathcal{K}=0$, thereby reducing the analysis to a fixed plane.

For a massive particle moving in the equatorial plane, the effective potential becomes
\begin{align}
V_{\text{eff}}(r) = \left(1 + \frac{\mathcal{L}^2  }{r^2} \right) {f(r)}.
\end{align}
\begin{figure}[ht]
    \centering
    \includegraphics[width=1.0\linewidth]{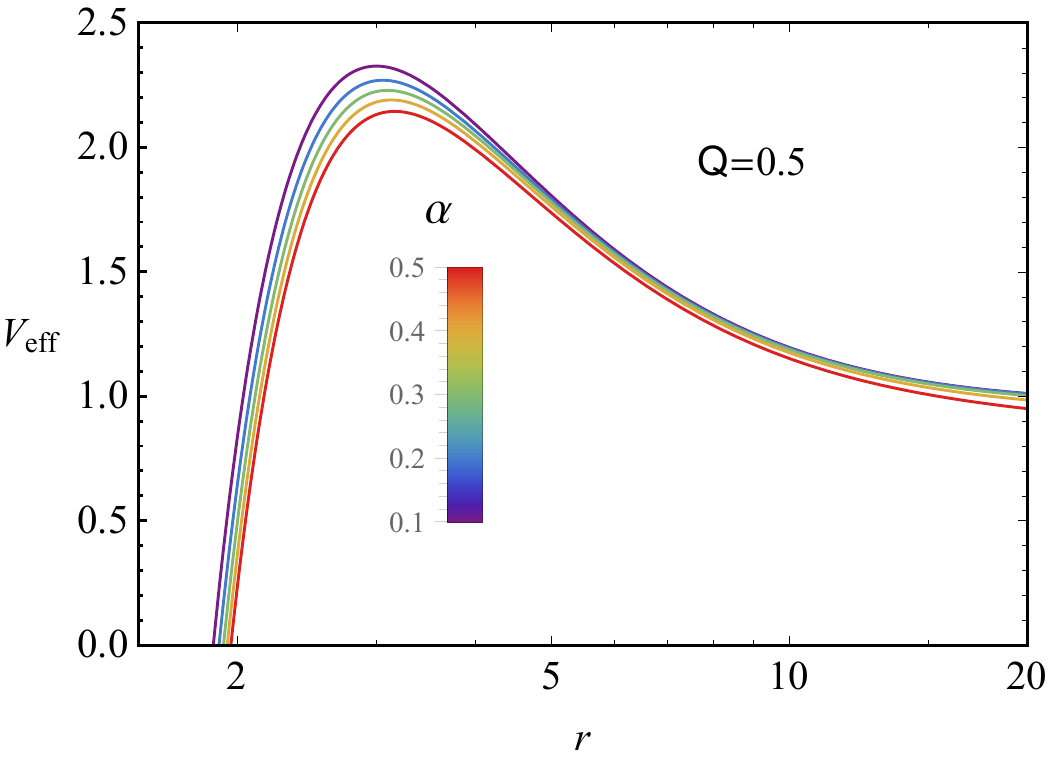}
    \caption{Effective potential $V_{\text{eff}}$ for massive particle with fixed $Q=0.5$ and varing Yukawa parameter $\alpha$.
    }
    \label{V1}
\end{figure}
\begin{figure}[ht]
    \centering
    \includegraphics[width=1.0\linewidth]{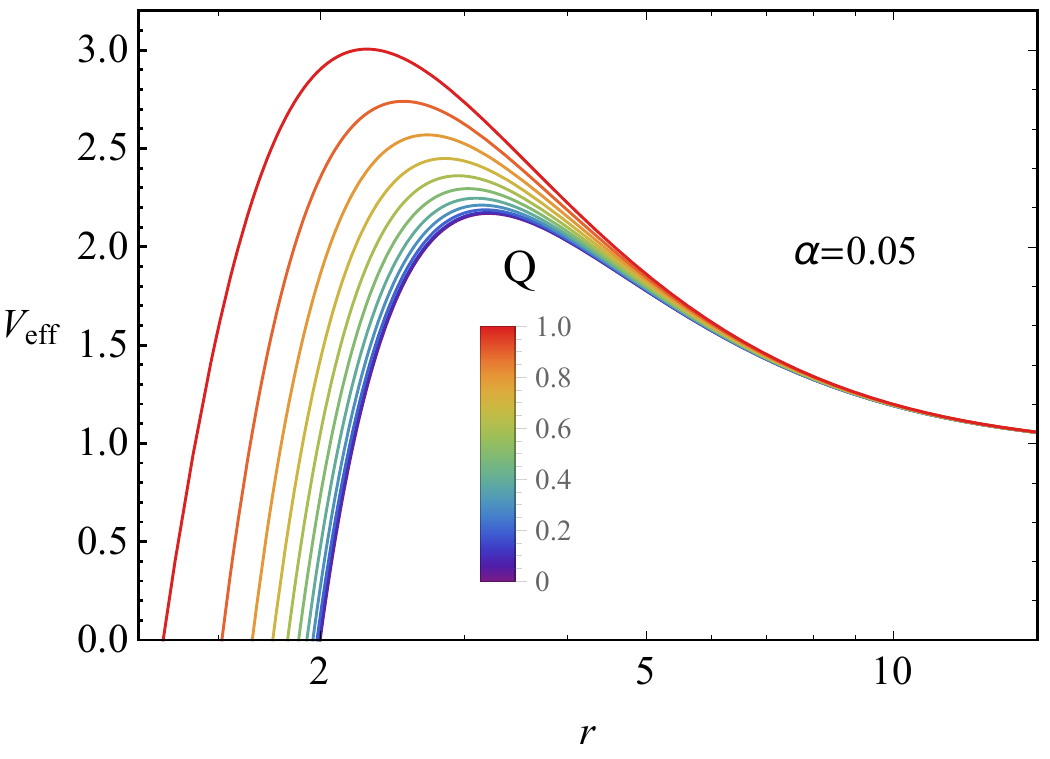}
    \caption{Effective potential $V_{\text{eff}}$ for massive particle with fixed $\alpha=0.05$ and varing charge parameter $Q$.
    }
    \label{V2}
\end{figure}
Figs.~\ref{V1} and~\ref{V2} illustrate how the charge parameter $Q$ and the Yukawa parameter $\alpha$ influence the dynamics of massive test particles around the ENLMY BH. The peak of the potential represents the unstable circular orbit, while the region where the potential exhibits a minimum corresponds to a stable circular motion. In Fig.~\ref{V1}, variations in $\alpha$ represent the influence of the Yukawa correction to gravity, which reshapes the potential profile and alters the radius of ISCO. In Fig.~\ref{V2}, varying $Q$ alters the strength of the electromagnetic contribution, modifying the potential barrier and shifting the location of unstable orbits. As $r\rightarrow\infty$, all gravitational, electromagnetic, and Yukawa contributions vanish, the spacetime becomes asymptotically flat, and the effective potential approaches the rest-mass level $V_{\text{eff}}\rightarrow1$, indicating that the particle behaves like a free particle far from the BH.

\subsection{Circular Orbits of the Particle Motion}
\begin{figure}[ht]
    \centering
    \includegraphics[width=1.0\linewidth]{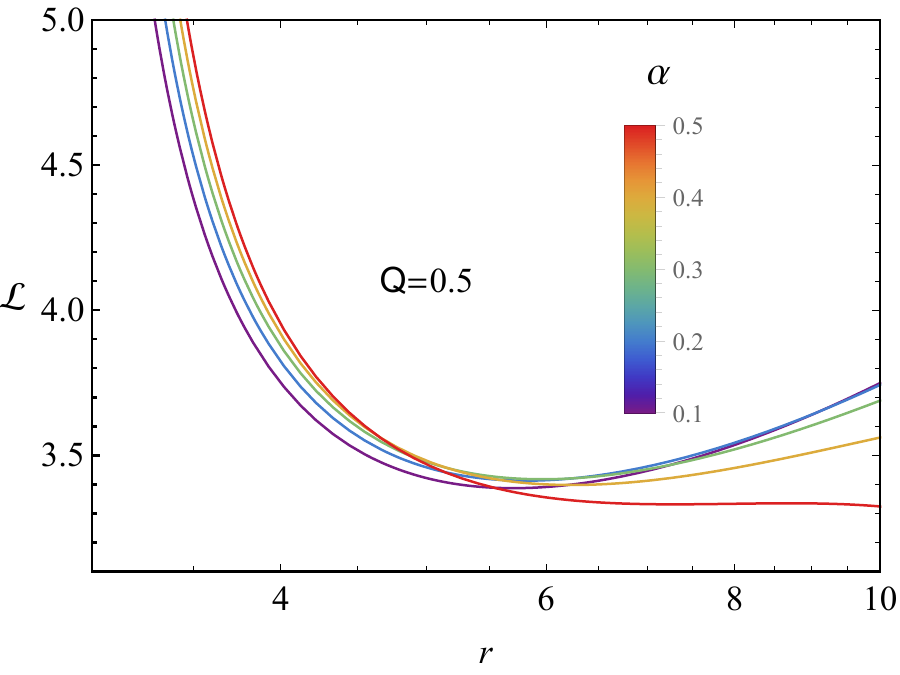}
    \caption{Variation of angular momentum $\mathcal{L}$ with radial coordinate $r$ for fixed $Q=0.5$ and varying $\alpha$.}
    \label{L1}
\end{figure}

\begin{figure}[ht]
    \centering
    \includegraphics[width=1.0\linewidth]{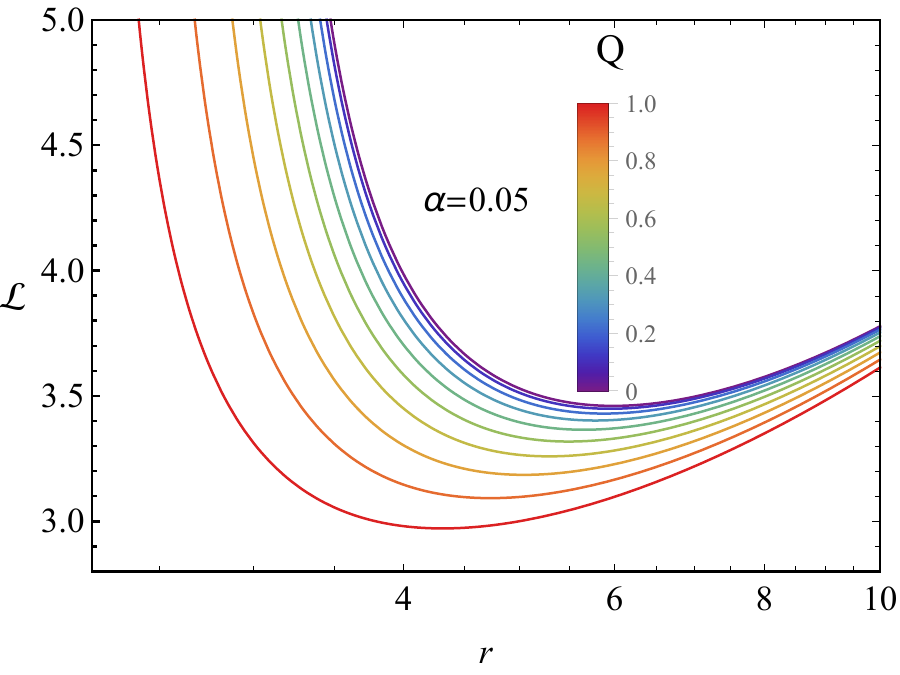}
    \caption{Variation of angular momentum $\mathcal{L}$ with radial coordinate $r$ for fixed $\alpha=0.05$ and varying $Q$.}
    \label{L2}
\end{figure}

The effective potential $V_{\text{eff}}(r)$ plays a crucial role in understanding the motion of test particles, as it provides insight into the nature of their trajectories. The extrema of $V_{\text{eff}}(r)$ correspond to circular orbits, where a minimum indicates a stable circular orbit and a maximum indicates an unstable one.
The particle moving in the circular orbits in the equatorial plane of ENLMY BH, satisfies the conditions \cite{2017bhlt.book.....B,2025PDU....4701816S}
\begin{align}
V_{\text{eff}}(r) = \mathcal{E}^2, \quad \frac{d V_{\text{ eff}} (r)}{d r} = 0.
\end{align}
From these conditions, the angular momentum $\mathcal{L}$ of the particle moving in the circular orbits can be calculated as
\begin{align}\label{ang}
    \mathcal{L}^2=\frac{r^2X_1}{X_2},
\end{align}
where
\begin{align}\label{X}
    X_1 =&12 Q^2-12e^{r\alpha}Mr+4Q^2r\alpha-2Q^2r^2\alpha^2+Q^2r^3\alpha^3 \nonumber\\
    &+e^{r\alpha}Q^2r^3\alpha^3-Q^2r^4\alpha^4+2e^{r\alpha}Q^2r^4\alpha^4\mathcal{E}_1(\alpha r),\\
    X_2=&-24Q^2+36e^{r\alpha} Mr-12e^{r\alpha}r^2-Q^2r^3\alpha^3\nonumber\\ &+e^{r\alpha}Q^2r^3\alpha^3+Q^2r^4\alpha^4,
\end{align}
while the energy of the particle comes out to be
\begin{align}\label{Eng}
\mathcal{E}^2=-\frac{e^{-r\alpha}Y_1^2}{3r^2Y_2},
\end{align}
where
\begin{align}\label{Y}
    Y_1=&-6Q^2+12e^{r\alpha}Mr-6e^{r\alpha}r^2+2Q^2r\alpha-Q^2r^2\alpha^2\nonumber\\
    &+e^{r\alpha}Q^2r^3\alpha^3+e^{r\alpha}Q^2r^4\alpha^4\mathcal{E}_1(\alpha r),\\
    Y_2=&-24Q^2+36e^{r\alpha}Mr-12e^{r\alpha}r^2-Q^2r^3\alpha^3\nonumber\\
    &+e^{r\alpha}
Q^2r^3\alpha^3+Q^2r^4\alpha^4\end{align}
Figs.~\ref{L1} and~\ref{L2} illustrate how the angular momentum $\mathcal{L}$ of a test particle orbiting the ENLMY BH depends on the radial distance $r$ for different values of charge $Q$ and the Yukawa parameter $\alpha$. In Fig.~\ref{L1}, increasing $\alpha$ modifies the effective gravitational potential, resulting in a shift in the angular momentum profile, indicating that higher $\alpha$ weakens the gravitational attraction at intermediate distances, thereby increasing the angular momentum needed to maintain a circular orbit. In Fig.~\ref{L2}, increasing $Q$ shifts the minimum of $\mathcal{L}$ slightly outward and changes the required angular momentum for stable orbits, reflecting the stronger electrostatic repulsion. 


\begin{figure}[ht]
    \centering
    \includegraphics[width=1.0\linewidth]{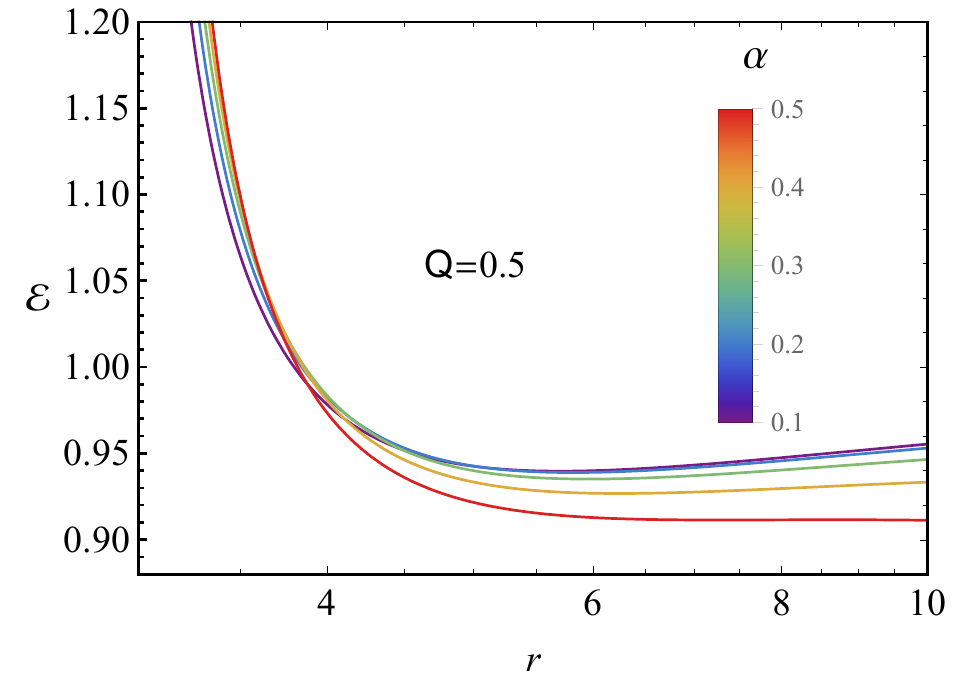}
    \caption{Variation of the energy $\mathcal{E}$ with radial coordinate $r$ for fixed $Q=0.5$ and varying Yukawa parameter $\alpha$.
    }
    \label{E1}
\end{figure}
\begin{figure}[ht]
    \centering
    \includegraphics[width=1.0\linewidth]{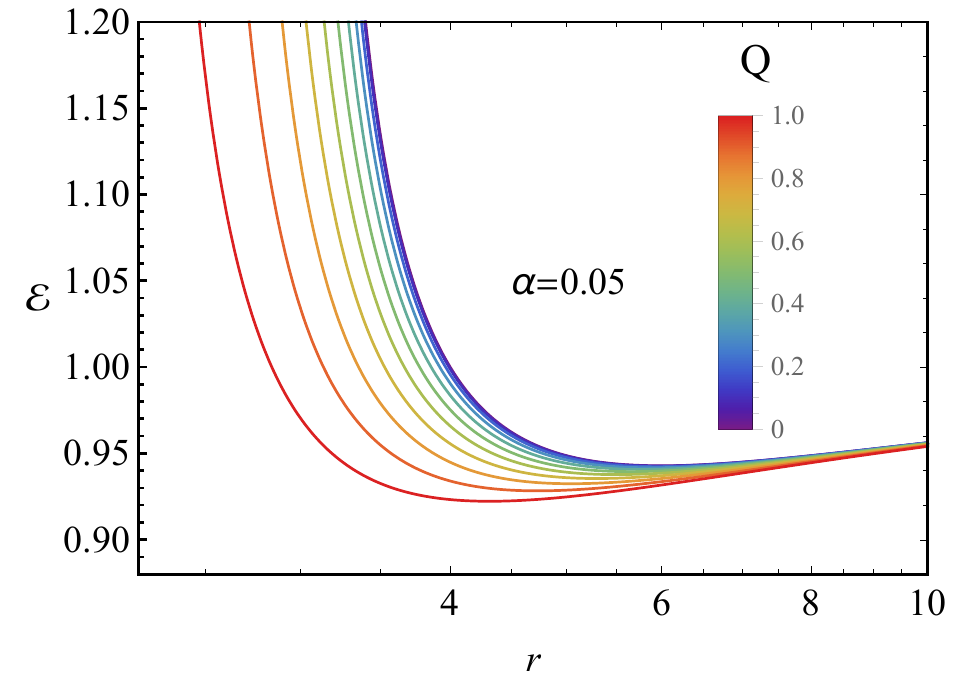}
    \caption{Variation of the energy $\mathcal{E}$ with radial coordinate $r$ for fixed $\alpha=0.05$ and varying charge $Q$.
    }
    \label{E2}
\end{figure}
In Fig.~\ref{E1}, for fixed electric charge $Q=0.5$, increasing the Yukawa parameter $\alpha$ slightly decreases the required energy at all radii. This indicates that a stronger Yukawa screening parameter weakens gravitational attraction, so particles require less energy to maintain an orbit at a given radius. In Fig.~\ref{E2}, where $\alpha=0.05$ is fixed, and the charge $Q$ is varied, increasing $Q$ weakens the effective gravitational attraction, so the particle needs lower energy to orbit at the same radius. This occurs because the charge introduces a repulsive effect, further reducing the gravitational pull and lowering the energy required for circular motion. In both plots, the sharp rise of $\mathcal{E}$ at small $r$ reflects the strong relativistic effects near the horizon, while at large $r$, the curves flatten, approaching the Newtonian limit.

\subsection{ISCO}
The minima and maxima of the effective potential correspond to the locations of stable and unstable circular orbits of a test particle. The shape of the effective potential depends on the particle’s angular momentum as well as the parameters of the BH. For compact objects such as a Schwarzschild BH, extrema in the effective potential appear only for specific values of the angular momentum. The minimum corresponds to a stable circular orbit, while the maximum corresponds to an unstable one. The ISCO for a Schwarzschild BH is located at $r=6M$ (equivalently $r=3r_g$, where $r_g=2M$ is the Schwarzschild radius). The ISCO of a particle orbiting a general compact object can be determined by solving \cite{2017bhlt.book.....B,2025PDU....4701816S}
\begin{align}
\frac{d^2   V_{\text{eff}} (r)} {dr^2} = 0,\label{ISCO}
\end{align}
which ensures the presence of an inflection point marking the transition between stable and unstable orbits.

Our investigation of ISCO for metric \eqref{eq.ds} shows consistent physical behavior across key parameter regimes. For uncharged BHs ($Q = 0$), we recover the Schwarzschild ISCO at $r = 6M$. In the Yukawa screening parameter limit $\alpha\rightarrow 0$, the ISCO radii smoothly approach RN values, mirroring the metric's reduction to the RN solution. For general cases with $\alpha > 0$ and $Q > 0$, we observe modified ISCO radii demonstrating the Yukawa correction's influence on orbital stability as compared to RN BHs.

\begin{figure}[ht]
    \centering
    \includegraphics[width=1.0\linewidth]{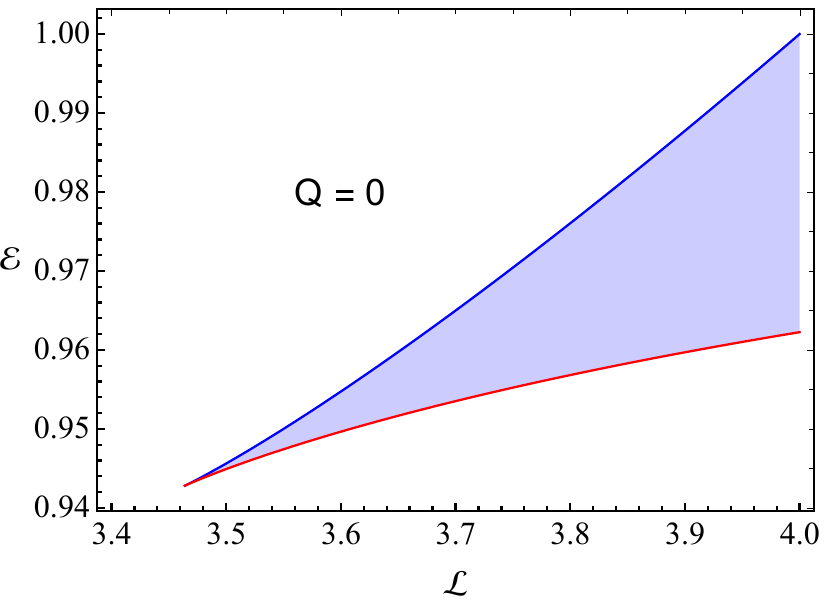}\caption{Allowed region (in the shadow) of the $(\mathcal{L},\mathcal{E})$ plane for $Q=0$ and $\alpha=0.05$ around the ENLMY BH. }\label{E-L1}
    \end{figure}
    \begin{figure}[ht]
    \centering
    \includegraphics[width=1.0\linewidth]{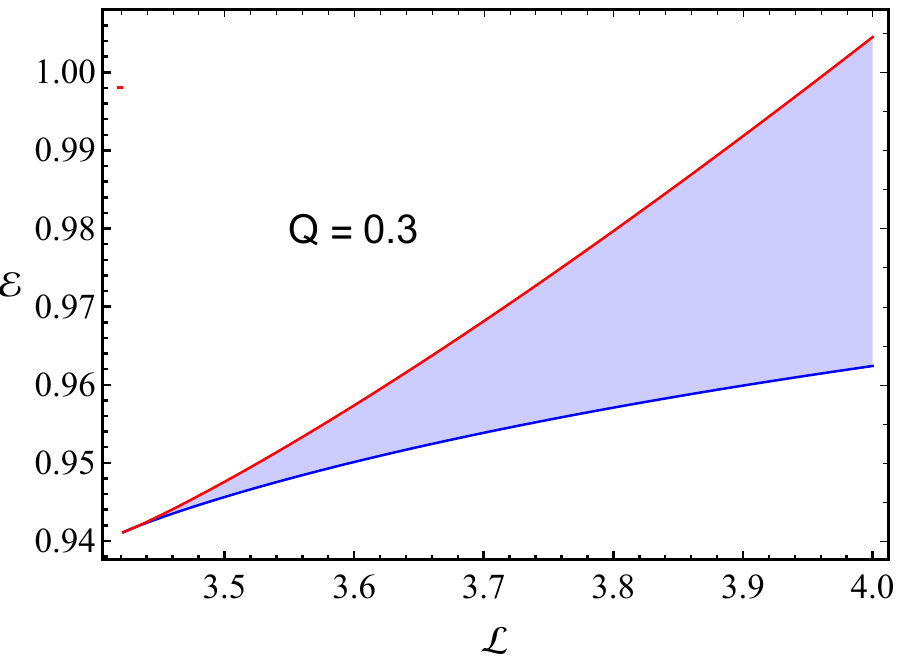}\caption{Allowed region (in the shadow) of the $(\mathcal{L},\mathcal{E})$ plane for $Q=0.3$ and $\alpha=0.05$ around the ENLMY BH.}\label{E-L2}
    \end{figure}
    \begin{figure}[ht]
    \centering
     \includegraphics[width=1.0\linewidth]{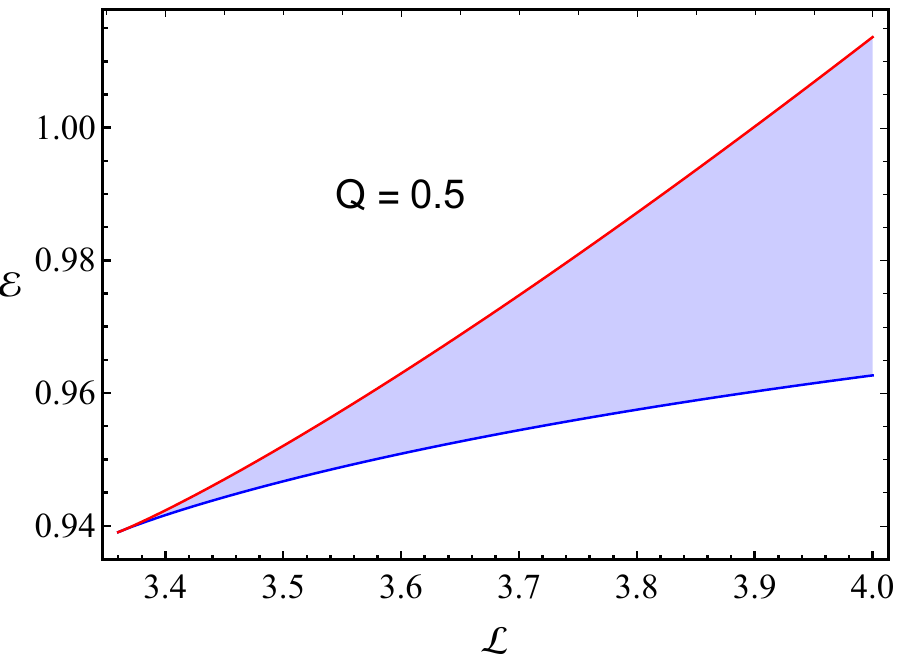}
    \caption{Allowed region (in the shadow) of the $(\mathcal{L},\mathcal{E})$ plane for $Q=0.5$ and $\alpha=0.05$ around the ENLMY BH.}\label{E-L3}
\end{figure}

The IBCO represents the smallest radius at which a particle can maintain a circular orbit with zero binding energy, meaning the total energy of the particle equals its rest-mass energy. At this threshold, the particle is marginally bound to the BH. For a Schwarzschild BH, the IBCO occurs at $r=4M$ (or $r=2r_g$, where $r_g=2M$ is the Schwarzschild radius). To determine the IBCO for a general spacetime, one imposes the conditions for circular motion \cite{2017bhlt.book.....B,2025PDU....4701816S}
\begin{align}\label{IBCO}
V_{\text{eff}}=\mathcal{E}^2=1,\quad\frac{d V_{\text{eff}} (r)} {dr} = 0.
\end{align}
Solving this equation simultaneously yields the radius at which the particle transitions from bound to unbound circular motion in the gravitational field of the compact object. Using Eqs.~\eqref{ISCO} and~\eqref{IBCO}, the radii of the ISCO and IBCO can be determined for the given spacetime. However, due to the complexity of the resulting equations, the conditions for circular motion and stability cannot be solved analytically. Therefore, we compute the ISCO and IBCO radii numerically by solving the corresponding equations for the effective potential and energy simultaneously. Figs.~\ref{E-L1}-\ref{E-L3} illustrate the allowed regions in the $\mathcal{E}-\mathcal{L}$ plane for bound orbits around the ENLMY BH, where $Q=0$ corresponds to the Schwarzschild spacetime.

\section{Periodic Orbits}
Periodic orbits provide a detailed understanding of particle motion around compact objects, representing trajectories in which a particle returns to its initial position after a finite time. In the background of an ENLMY BH, such orbits occur when the ratio of the radial $\omega_r$ and angular $\omega_\phi$ frequencies is a rational number, ensuring that the motion repeats after a certain number of cycles. To systematically classify these orbits, we use a rational number $q$ defined in terms of orbital frequencies, along with a triplet of integers $(z,w,v)$, where $z$ denotes the number of radial oscillations (zooms), $w$ represents the number of complete rotations (whirls) around the BH, and $v$ counts the number of vertices or angular turning points per radial period. The rational number characterizing the periodicity can be written as \cite{2009PhRvL.103m1101H,2025PDU....4701816S}
\begin{align}\label{rational}
    q=w+\frac{v}{z}=\frac{\omega_\phi}{\omega_r}-1=\frac{\Delta\phi}{2\pi}-1.
\end{align}
To prevent orbital degeneracy, the integers $z$ and $v$ must be relatively prime. The allowed values of $v$ are then $v=
   \begin{cases}
    1\leq v\leq z-1,\qquad &\text{if}~ z>1,\\
    0,\qquad&\text{if}~ z=1
    \end{cases}$ \cite{2008PhRvD..77j3005L}.
The motion in the $\phi$ and $r$ directions is governed by the angular $\omega_\phi$ and radial $\omega_r$ frequencies, respectively, which are defined as follows
\begin{align}\label{frEqs.}
    \omega_r=\frac{2\pi}{T_r},\quad\omega_\phi=\frac{1}{T_r}\int_{0}^{T_r}\frac{d\phi}{dt}dt=\frac{\Delta\phi}{T_r},
\end{align}
where $T_r$ indicates the (coordinate) time for one full radial oscillation of the particle, and $\Delta\phi$ denotes the total accumulated angle during that period. The expression for $\Delta\phi$ is given by
\begin{align}\label{DeltaPhi}
    \Delta\phi = 2 \int_{r_1}^{r_2} \frac{\dot{\phi}}{\dot{r}}\,dr 
= 2 \int_{r_1}^{r_2}\frac{\mathcal{L}}{r^2\sqrt{{\mathcal{E}}^2-\left(1+\frac{\mathcal{L}^2}{r^2}\right)f(r)}}dr.
\end{align}
Here, $r_1$ and $r_2$ denote the two turning points of the orbit between the ISCO and IBCO, commonly referred to as the apastron and periastron, respectively. Using Eqs.~\eqref{rational} and~\eqref{DeltaPhi}, we construct the Tables~\ref{tab1} and~\ref{tab2} listing the energy and angular momentum values corresponding to different 
$(z,w,v)$. In these tables, the angular momentum is fixed as $\mathcal{L}_{\text{av}}=({\mathcal{L}_{\text{ISCO}}+\mathcal{L}_{\text{IBCO}}})/{2}$, where $\mathcal{L}_{\text{av}}$ depends on the parameters $\alpha$, $M$, and $Q$. The angular momentum of the particle satisfies the 
$\mathcal{L}_{\text{ISCO}}<\mathcal{L}_{\text{av}}<\mathcal{L}_{\text{IBCO}}$ \cite{2008PhRvD..77j3005L,2025PDU....4701816S}.

\begin{table*}
\centering
    \caption{Energy and angular momentum of the particle corresponding to the periodic orbits around the ENLMY BH with values of the parameters $z=1,2,3,w=1,v=1$, $Q=0.5$ and $\alpha=0.1,0.2,0.3$.}
     \label{tab1}
    \begin{tabular}{cccccc}
    \toprule
      $\alpha$ & ${\cal L}_{\text{av}}$ &${\cal E}_{(1,1,0)}$  &${\cal E}_{(2,1,1)}$  & ${\cal E}_{(3,1,1)}$ & ${\cal E}_{(4,1,1)}$\\
      \midrule
      0.1&  3.65484 &  0.959765& 0.958564 &  0.959891    & 0.959835  \\
        0.2 &  3.6881 &  0.959797  & 0.959255 & 0.959368 &  0.959355 \\
         0.3 &  3.71326 & 0.948468 & 0.949613   &  0.948305  & 0.949752 \\
         \bottomrule
    \end{tabular}
    \end{table*}

\begin{table*}
\centering
    \caption{Energy values and corresponding angular momentum for periodic orbits with parameters $(z=1,2,3,w=1,v=1)$ around the ENLMY BH for $\alpha=0.05$ and various values of $Q=0,0.1,0.2,0.3,0.4,0.5$. Similar results are obtained for $Q$ in the range $0.6\leq Q\leq 1.0$.}
     \label{tab2}
    \begin{tabular}{cccccc}
    \toprule
      $Q$ & ${\cal L}_{\text{av}}$ &${\cal E}_{(1,1,0)}$  &${\cal E}_{(2,1,1)}$  & ${\cal E}_{(3,1,1)}$ & ${\cal E}_{(4,1,1)}$\\
      \midrule
      0&  3.73205 &0.965426  & 0.968027 & 0.967645 &0.967334  \\
      0.1&  3.72824 & 0.959689 & 0.959824 &  0.958046    & 0.959821  \\
        0.2 &  3.71673 &  0.959781  & 0.956952 & 0.959765 &  0.959933 \\
         0.3 &  3.69724 & 0.95981 &  0.957376  &  0.958707  & 0.959335\\
         0.4&  3.66926 & 0.959978  &  0.957926 &  0.959242  & 0.959919\\
         0.5&  3.63202 &  0.958983 & 0.95988 & 0.959905  & 0.959924 \\
         \bottomrule
    \end{tabular}
    \end{table*}
 \begin{figure*}
   \centering
\includegraphics[width=0.3\linewidth]{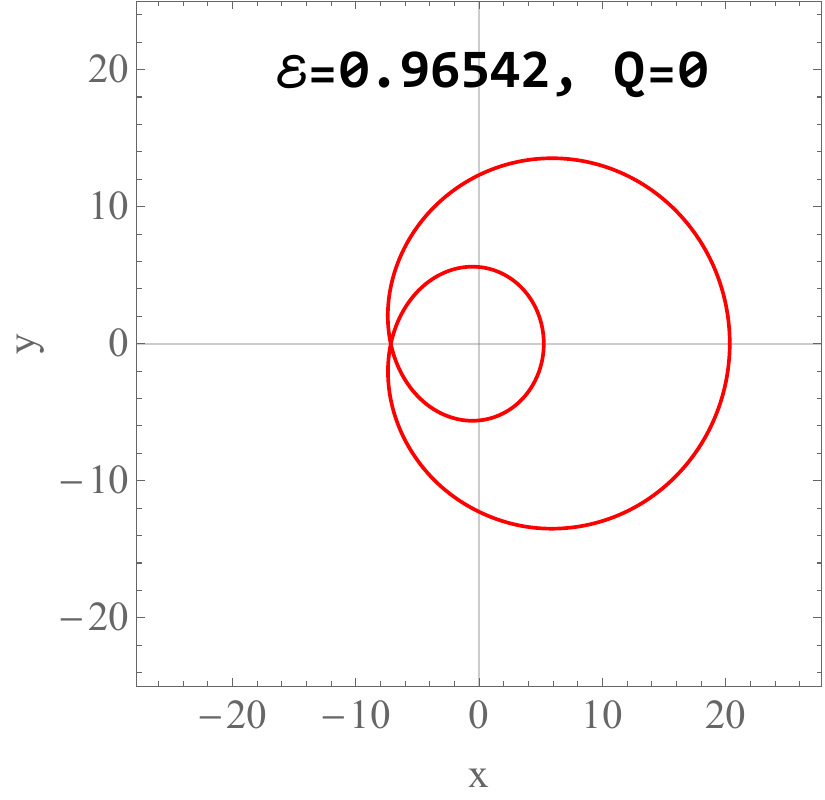}
\includegraphics[width=0.3\linewidth]{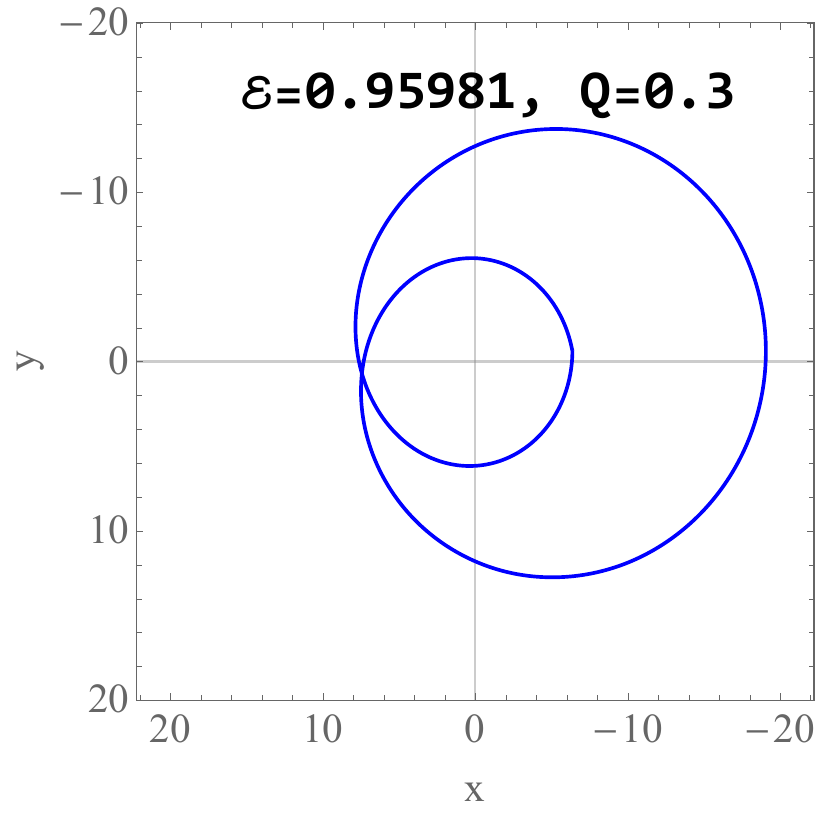}
\includegraphics[width=0.3\linewidth]{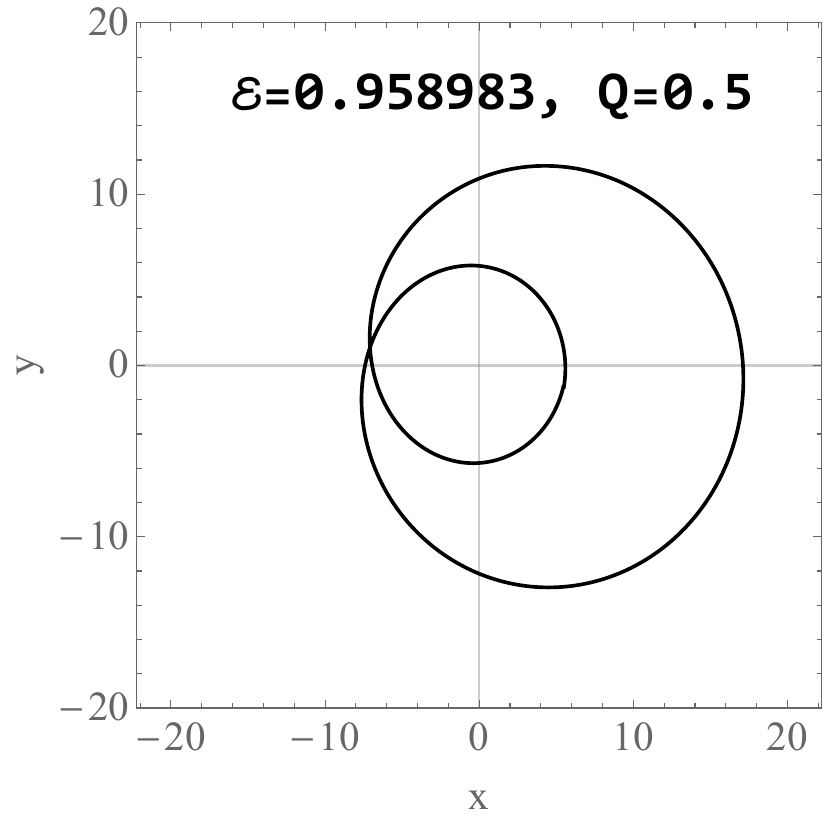}
\caption{ Zoom–whirl periodic orbits for the parameter set $(z,w,v)=(1,1,0)$ with $M=1$ and $\alpha=0.05$, plotted for $Q=0,0.3$, and $0.5$. The case $Q=0$ corresponds to the Schwarzschild spacetime. The trajectories are shown in the Cartesian plane using  $x={r}\cos\phi$  and $y={r}\sin\phi$. } \label{p1}
\end{figure*}
 \begin{figure*}
   \centering
\includegraphics[width=0.3\linewidth]{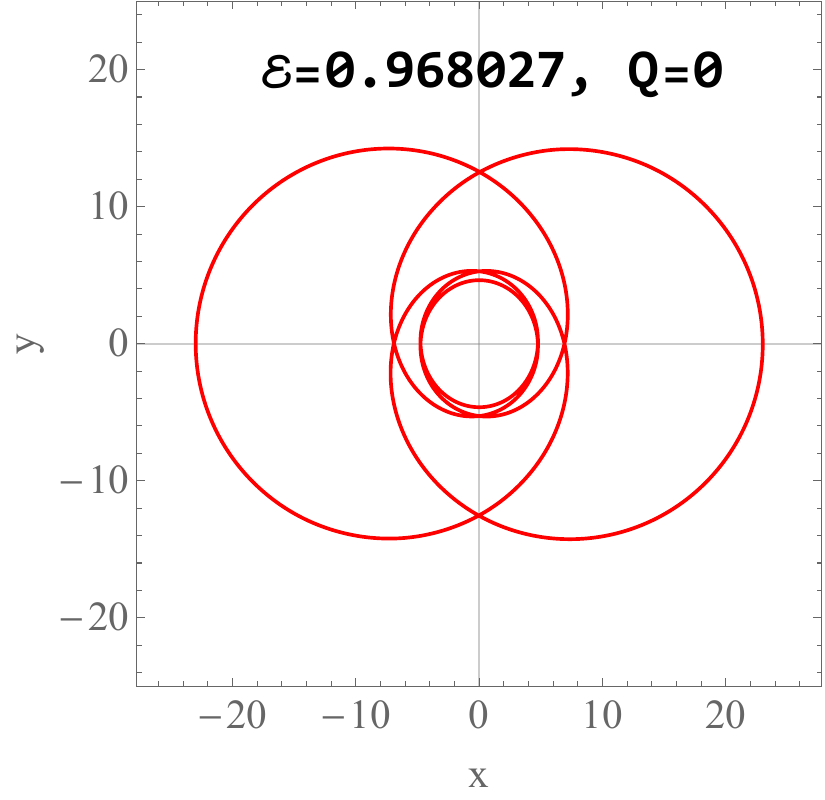}
\includegraphics[width=0.3\linewidth]{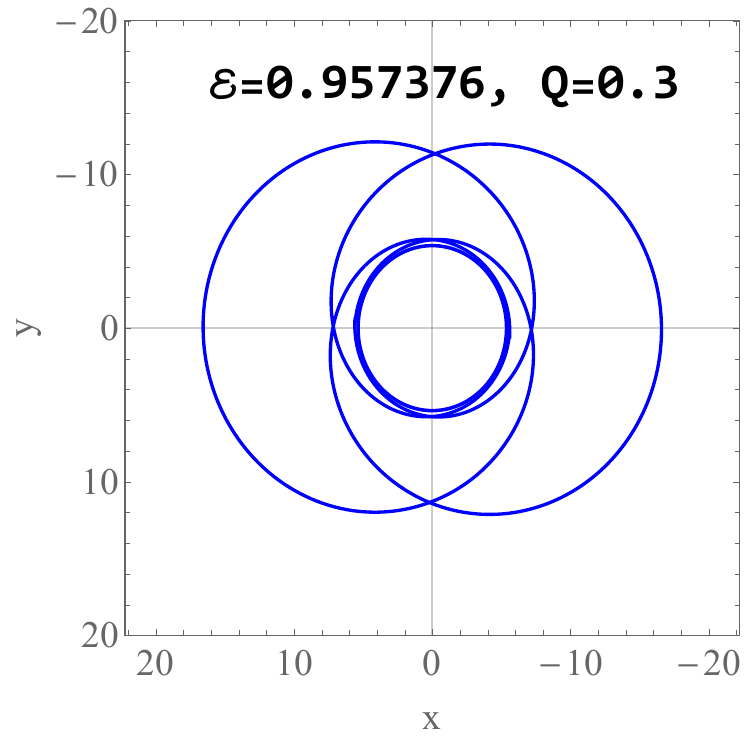}
\includegraphics[width=0.3\linewidth]{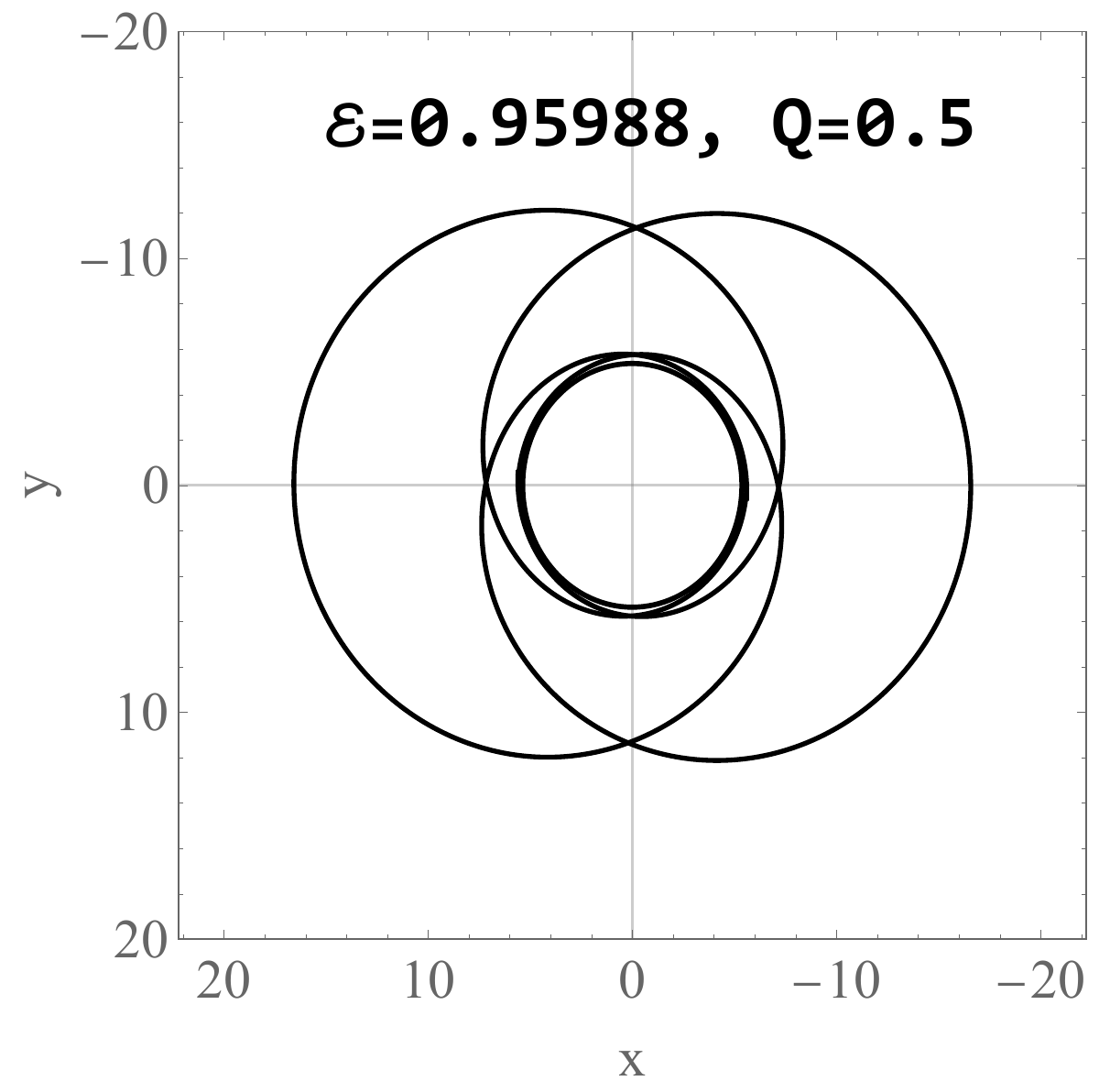}
\caption{ Zoom–whirl periodic orbits for the parameter set $(z,w,v)=(2,1,1)$ with $M=1$ and $\alpha=0.05$, plotted for $Q=0,0.3$, and $0.5$. The case $Q=0$ corresponds to the Schwarzschild spacetime. The trajectories are shown in the Cartesian plane using  $x={r}\cos\phi$  and $y={r}\sin\phi$. } \label{p2}
\end{figure*}
\begin{figure*}
   \centering
\includegraphics[width=0.3\linewidth]{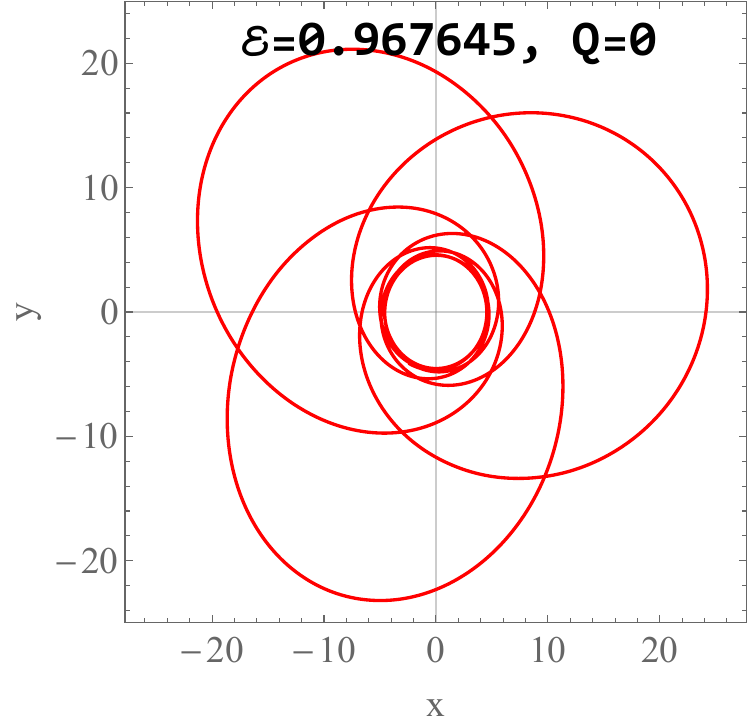}
\includegraphics[width=0.3\linewidth]{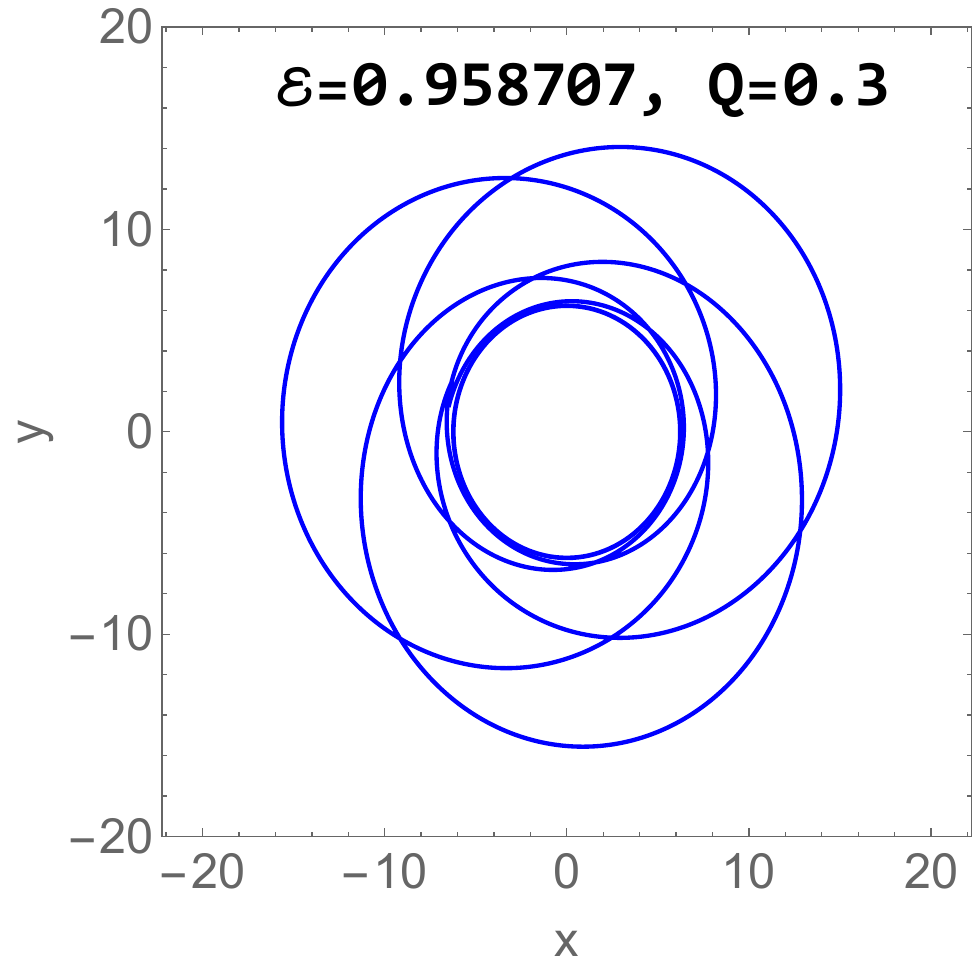}
\includegraphics[width=0.3\linewidth]{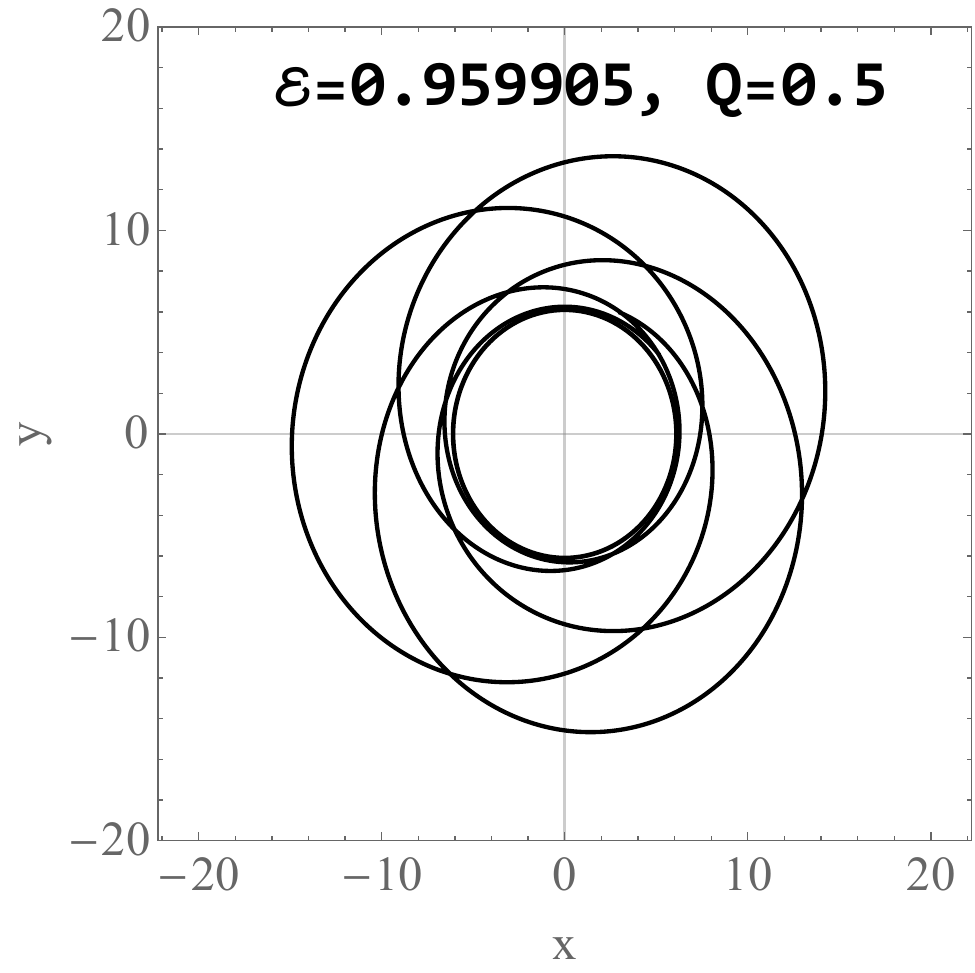}
\caption{ Zoom–whirl periodic orbits for the parameter set $(z,w,v)=(3,1,1)$ with $M=1$ and $\alpha=0.05$, plotted for $Q=0,0.3$, and $0.5$. The case $Q=0$ corresponds to the Schwarzschild spacetime. The trajectories are shown in the Cartesian plane using  $x={r}\cos\phi$  and $y={r}\sin\phi$. } \label{p3}
\end{figure*}

\begin{figure*}
   \centering
\includegraphics[width=0.3\linewidth]{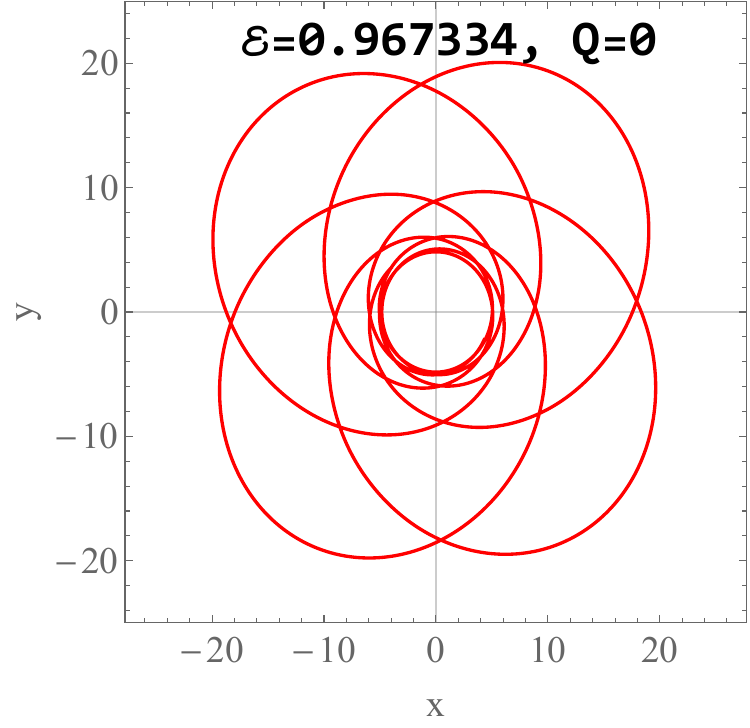}
\includegraphics[width=0.3\linewidth]{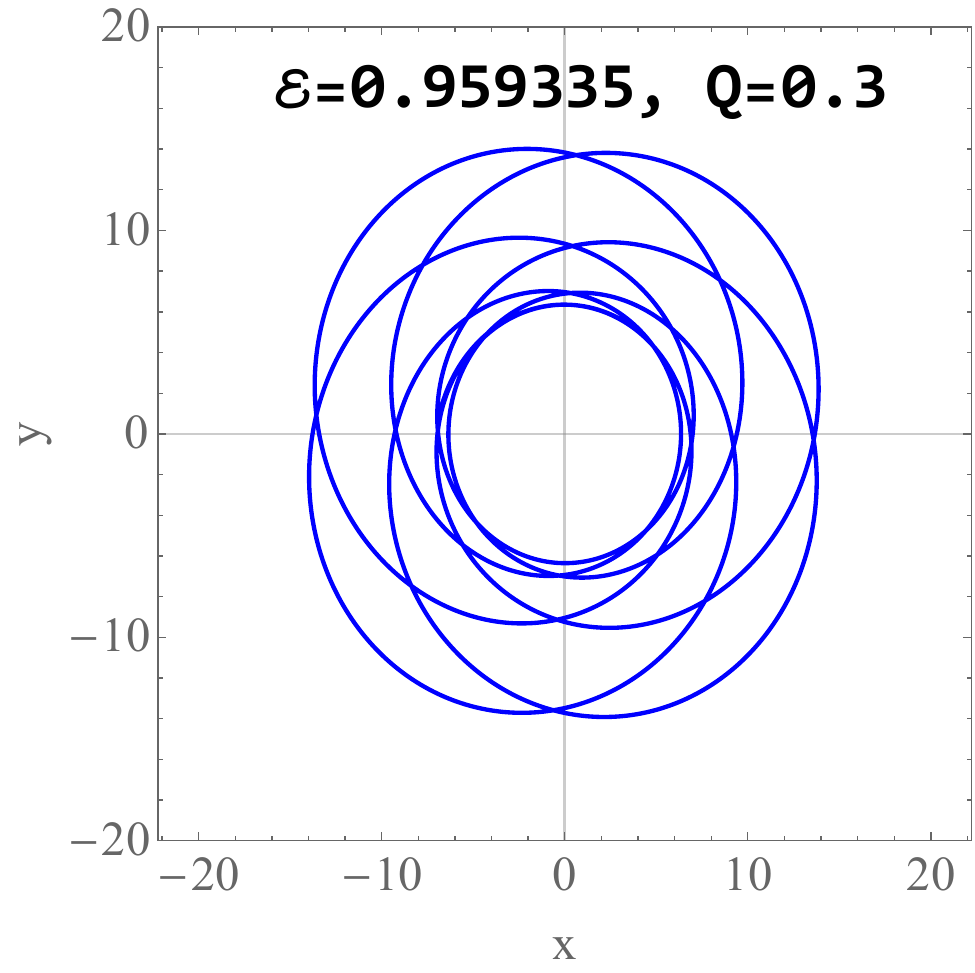}
\includegraphics[width=0.3\linewidth]{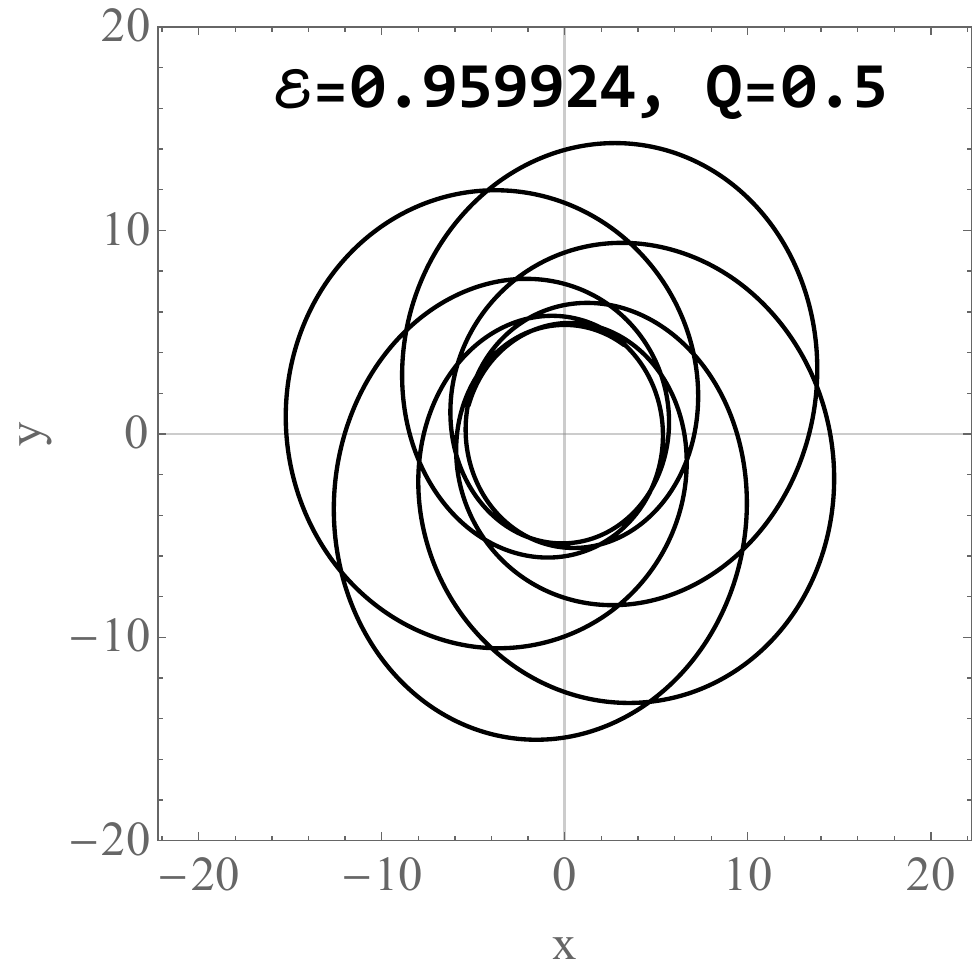}
\caption{ Zoom–whirl periodic orbits for the parameter set $(z,w,v)=(4,1,1)$ with $M=1$ and $\alpha=0.05$, plotted for $Q=0,0.3$, and $0.5$. The case $Q=0$ corresponds to the Schwarzschild spacetime. The trajectories are shown in the Cartesian plane using  $x={r}\cos\phi$  and $y={r}\sin\phi$. }\label{p4}
\end{figure*}

Figs.~\ref{p1}-\ref{p4} display the periodic orbits of massive particles between ISCO and IBCO in ENLMY spacetime for various values of $Q$ with $\alpha=0.05$. Each periodic orbit is characterized by a distinct triplet of integers, which corresponds to a unique energy value. Our comprehensive study reveals that the qualitative features of periodic orbits in the ENLMY spacetime remain broadly similar to those in the Schwarzschild spacetime. However, we observe that obtaining periodic orbits in the Schwarzschild spacetime generally requires higher energy compared to the ENLMY case. A similar result is obtained in \cite{2025EPJC...85.1340Z}. In both ENLMY spacetimes, the periodic orbit structures and energy patterns are comparable to those in the Schwarzschild case and require lower energy. In this paper, we initially considered values of $\alpha$ between 0.1 and 0.5. However, for higher values ($\alpha=0.4$ and $\alpha=0.5$), no periodic orbits are observed, as no region between the ISCO and IBCO can support such trajectories. This indicates that increasing $\alpha$ limits the range of allowed energies and radii, thereby preventing the formation of stable closed orbits at these higher values.

\section{Gravitational Wave Radiation from Periodic Orbits}
In this section, we present the analysis of the gravitational radiation emitted by periodic orbits of a test particle around the ENLMY BH. We model the system as an extreme-mass-ratio inspiral (EMRI), in which the mass of the smaller compact object is negligible compared to that of the central supermassive BH. This allows us to treat the smaller object as a perturbation to the ENLMY BH. Within this treatment, when the radiative losses of energy $\mathcal{E}$ and angular momentum $\mathcal{L}$ per orbital cycle remain sufficiently small, the adiabatic approximation becomes valid. Under this approximation, the particle's motion closely follows geodesic trajectories over several orbital periods, allowing us to track its periodic orbits and compute the associated gravitational radiation.
\par
To compute the GWs emitted by periodic orbits in the supermassive ENLMY BH spacetime, we adopt the kludge waveform approach introduced in \cite{2007PhRvD..75b4005B}. In this method, the small compact object is modeled as a test particle, and its periodic trajectory is obtained by solving the geodesic equations of the background spacetime. Once the orbital motion is determined, the corresponding GW signal is constructed using the standard quadrupole formula. The periodic orbits of a test particle in the ENLMY BH, derived in the preceding section by solving the geodesic equations, serve as the foundation for our GW analysis. The corresponding gravitational radiation emitted by these orbits can then be computed using the expression \cite{2022NatAs...6..464M,2023PhRvD.107d4053L}
\begin{align}\label{h}
    h_{ij}=\frac{4\eta M}{D_L}\left(V_i V_j-\frac{m}{r}n_i n_j\right),
\end{align}
evaluated up to the quadrupole approximation. Here, $M$ denotes the mass of ENLMY BH, $m$ is the mass of the test particle, and $D_L$ represents the luminosity distance from the EMRI system. The parameter $\eta=\frac{Mm}{\left(M+m\right)^2}$ is the symmetric mass ratio. Moreover, $V_i$ denotes the spatial velocity components of the test particle, and $n_i$ is the unit vector pointing along the radial direction associated with the test particle’s motion.

The GW signal can then be projected onto the detector-adapted coordinate frame, from which the two polarization modes, namely the plus $(h_+)$ and cross $(h_\times)$ components, are given as \cite{2022NatAs...6..464M,2023PhRvD.107d4053L,2025PDU....4701816S,2025EPJC...85.1340Z}

\begin{align}
h_+ &=-\frac{2\eta}{D_L}\frac{M^2}{r}\left(1+\cos^2\iota\right)\cos\left(2\phi+2\zeta\right),\\
    h_\times &=-\frac{4\eta}{D_L}\frac{M^2}{r}\cos \iota\sin\left(2\phi+2\zeta\right),
    \end{align}
    where $\iota$ denotes the inclination angle between the orbital angular momentum of the EMRI and the line of sight, and $\zeta$ is the latitudinal angle. To illustrate the GW waveforms associated with different periodic orbits and to assess the impact of ENLMY effects, we consider an EMRI system in which the secondary object has mass $m\ll M$. For simplicity, we fix the inclination angle $\iota$ and the latitudinal angle $\zeta$ to $\frac{\pi}{4}$. Moreover, the luminosity distance is taken to be a constant, $D_L=200 \text{Mpc}$. Under these assumptions, we arrive at the following conclusions,
    \begin{align}\label{h1}
    h_+ &\propto -\frac{\cos(2\phi+2\zeta)}{r},\\
h_\times &\propto -\frac{\sin(2\phi+2\zeta)}{r}.
\end{align}

    \begin{figure*}
   \centering
\includegraphics[width=0.3\linewidth]{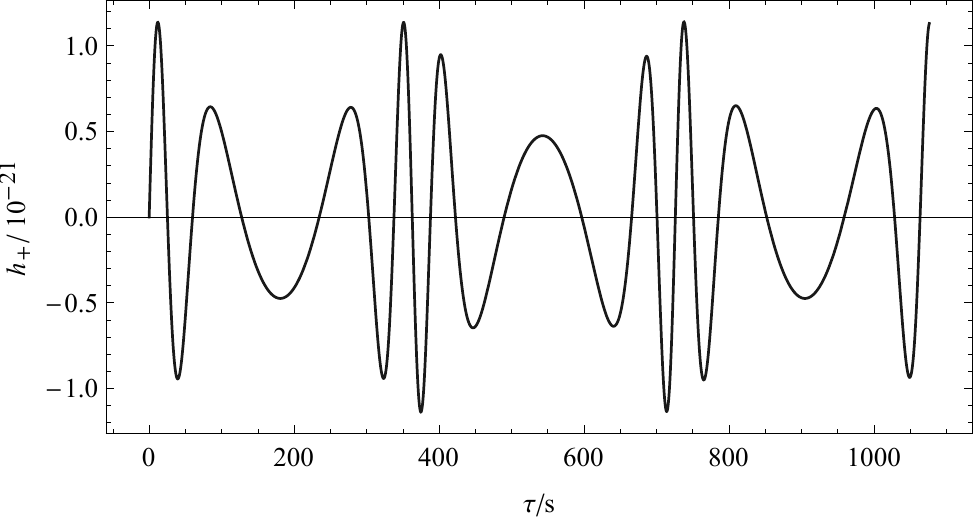}
\includegraphics[width=0.3\linewidth]{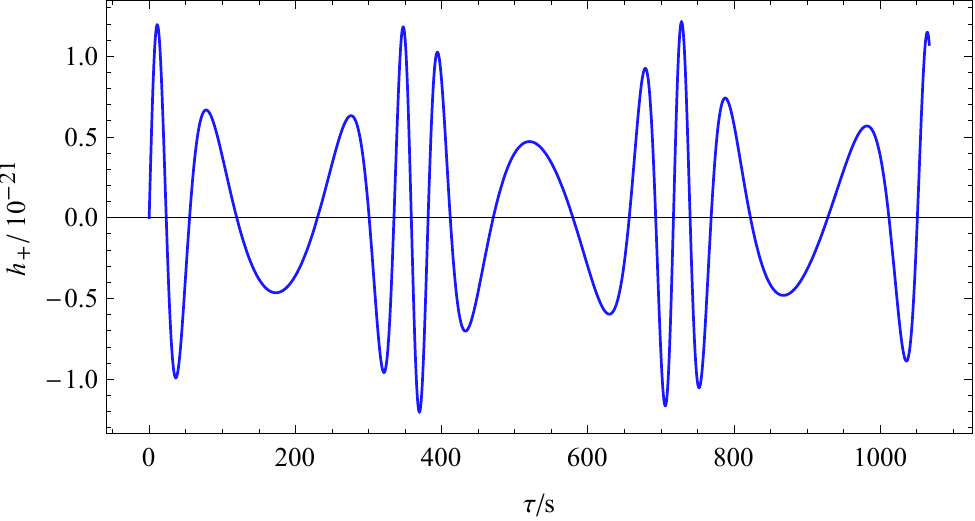}
\includegraphics[width=0.3\linewidth]{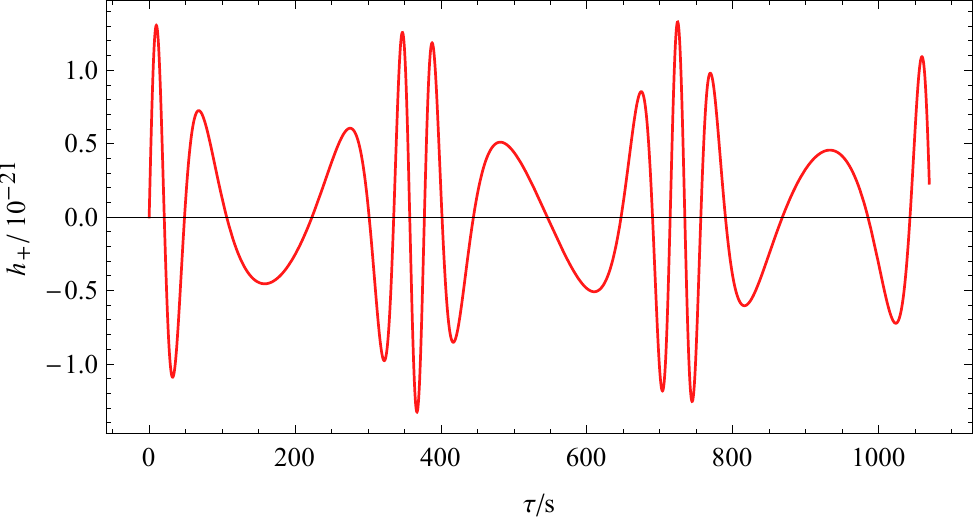} \caption{The $h_+$ polarization of the GW signal generated in the ENLMY BH for the periodic orbit characterized by the integer triplet $(3,1,1)$ with $\alpha=0.05$. The black line corresponds to $Q=0$ (Schwarzschild case), while the blue and red lines represent $Q=0.3$ and $Q=0.5$, respectively.}\label{h+311}
\end{figure*}

    \begin{figure*}[ht!]
   \centering
\includegraphics[width=0.3\linewidth]{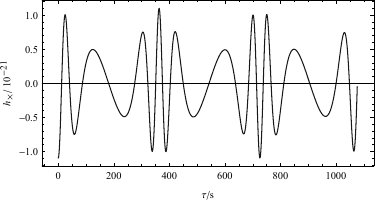}
\includegraphics[width=0.3\linewidth]{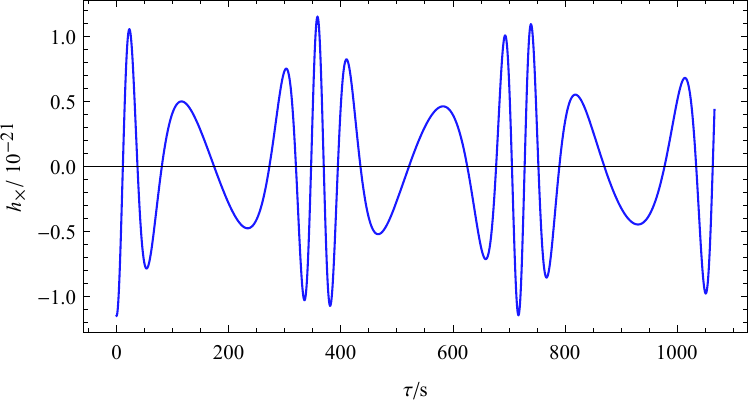}
\includegraphics[width=0.3\linewidth]{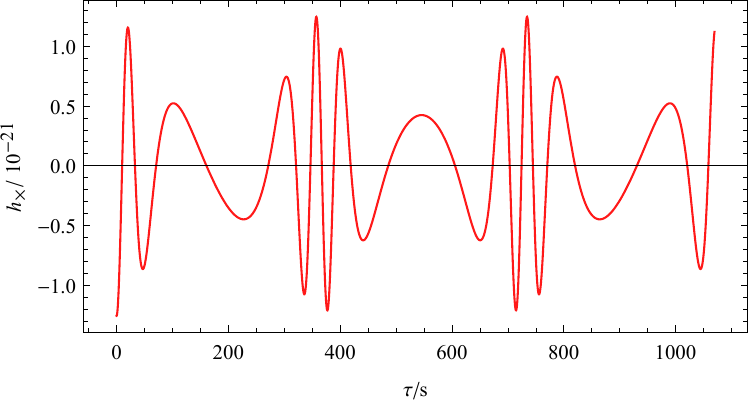} \caption{The $h_{\times}$ polarization of the GW signal generated in the ENLMY BH for the periodic orbit, characterized by the integer triplet $(3,1,1)$ with $\alpha=0.05$. The black line corresponds to $Q=0$ (Schwarzschild case), while the blue and red lines represent $Q=0.3$ and $Q=0.5$, respectively.}\label{h*311}
\end{figure*}

   \begin{figure*}[ht!]
   \centering
\includegraphics[width=0.45\linewidth]{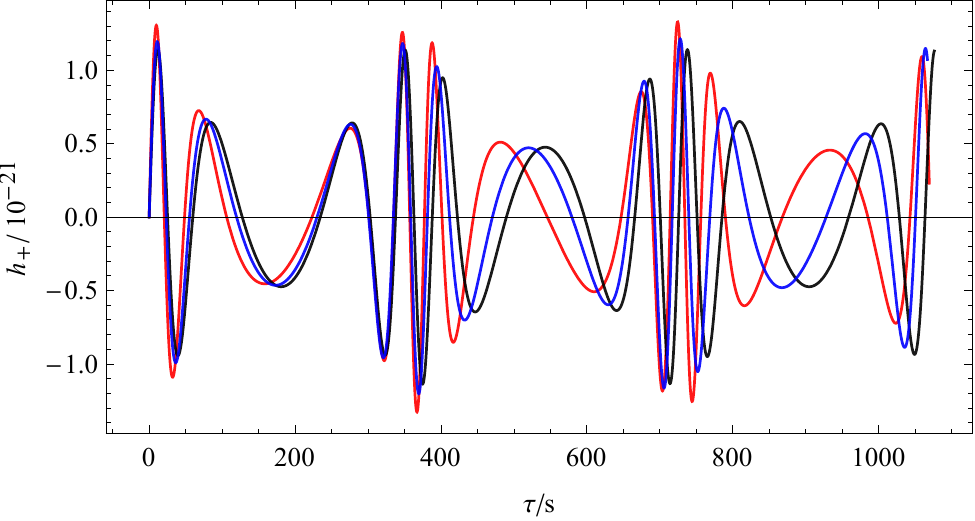}
\includegraphics[width=0.45\linewidth]{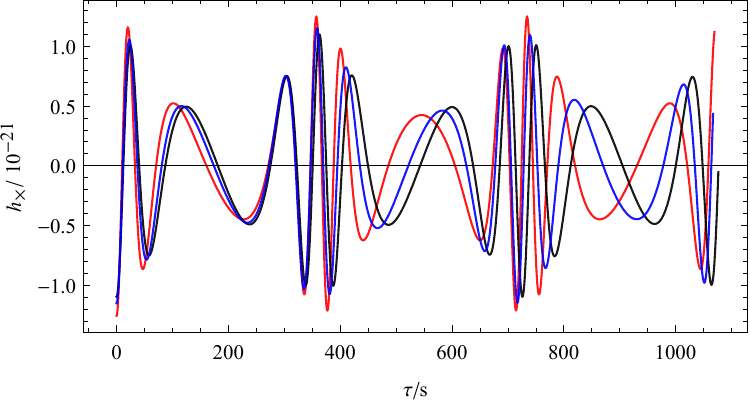}\caption{Right panel: Combined $h_+$ waveforms from Fig.~\ref{h+311}. Left panel: Combined $h_{\times}$ waveforms from Fig.~\ref{h*311}.}\label{h311combine}
\end{figure*}

\begin{figure*}
   \centering
\includegraphics[width=0.3\linewidth]{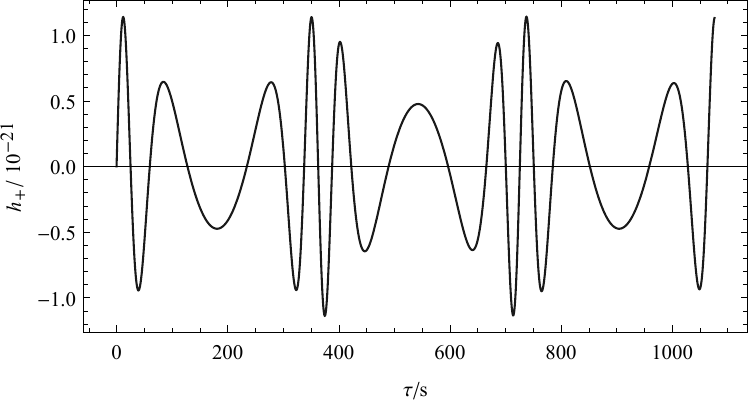}
\includegraphics[width=0.3\linewidth]{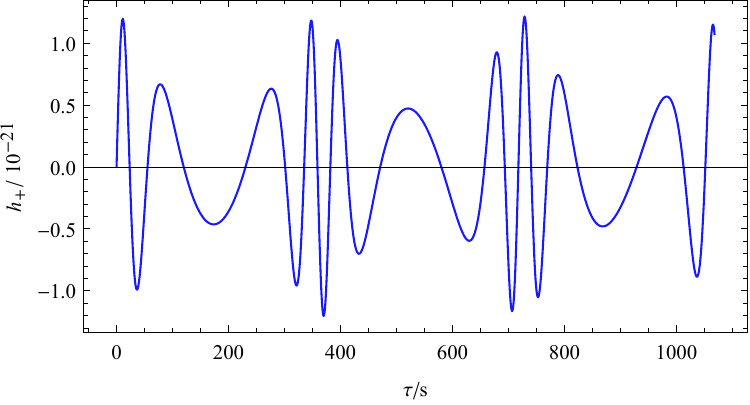}
\includegraphics[width=0.3\linewidth]{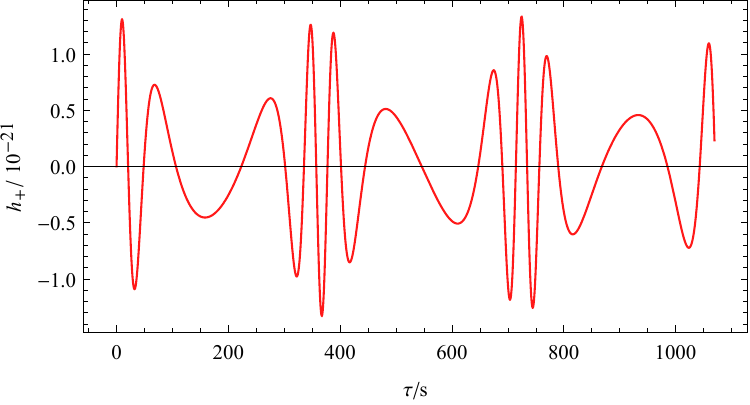} \caption{The $h_+$ polarization of the GW signal generated in the ENLMY BH for the periodic orbit characterized by the integer triplet $(4,1,1)$ with $\alpha=0.05$. The black line corresponds to $Q=0$ (Schwarzschild case), while the blue and red lines represent $Q=0.3$ and $Q=0.5$, respectively.}\label{h+411}
\end{figure*}

\begin{figure*}[ht!]
   \centering
\includegraphics[width=0.3\linewidth]{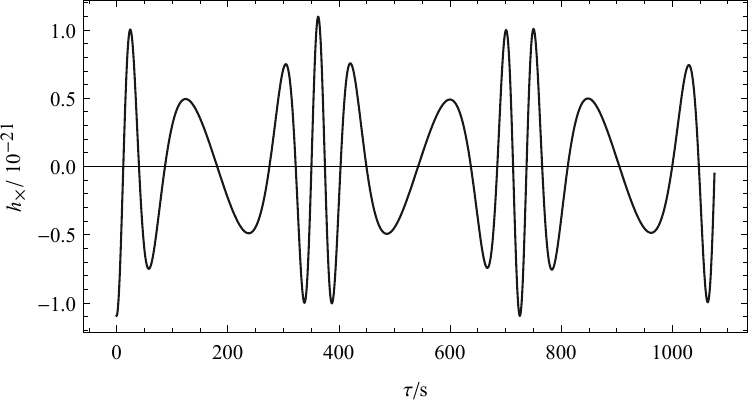}
\includegraphics[width=0.3\linewidth]{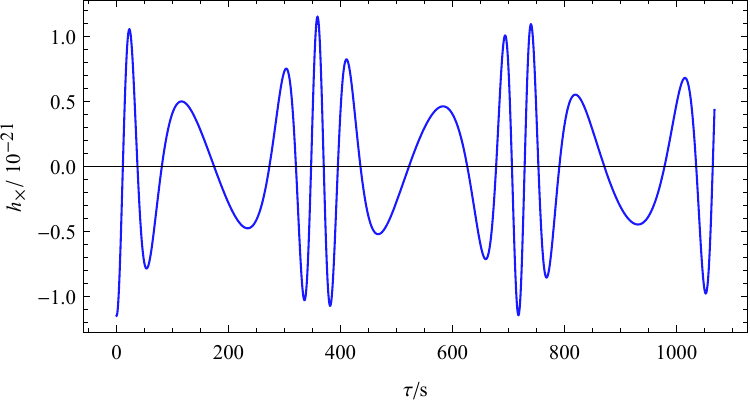}
\includegraphics[width=0.3\linewidth]{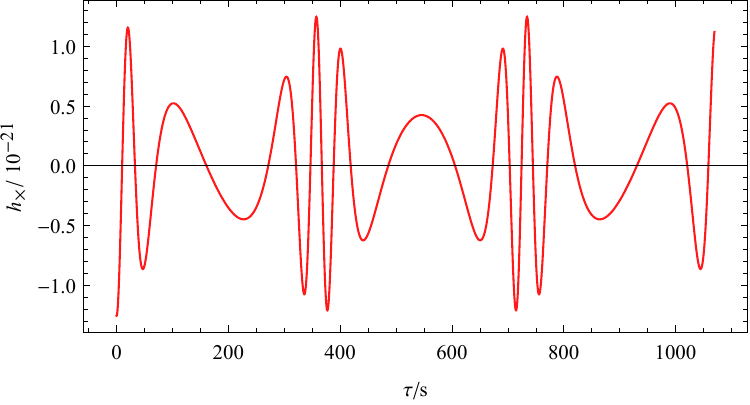} \caption{The $h_{\times}$ polarization of the GW signal generated in the ENLMY BH for the periodic orbit characterized by the integer triplet $(4,1,1)$ with $\alpha=0.05$. The black line corresponds to $Q=0$ (Schwarzschild case), while the blue and red lines represent $Q=0.3$ and $Q=0.5$, respectively.}\label{h*411}
\end{figure*}

\begin{figure*}[ht!]
   \centering
\includegraphics[width=0.45\linewidth]{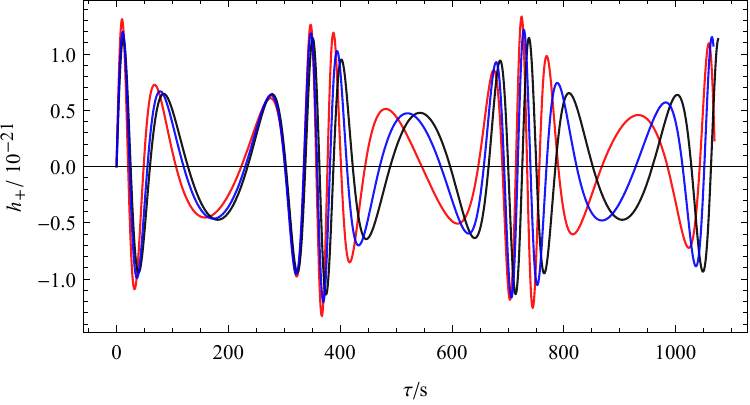}
\includegraphics[width=0.45\linewidth]{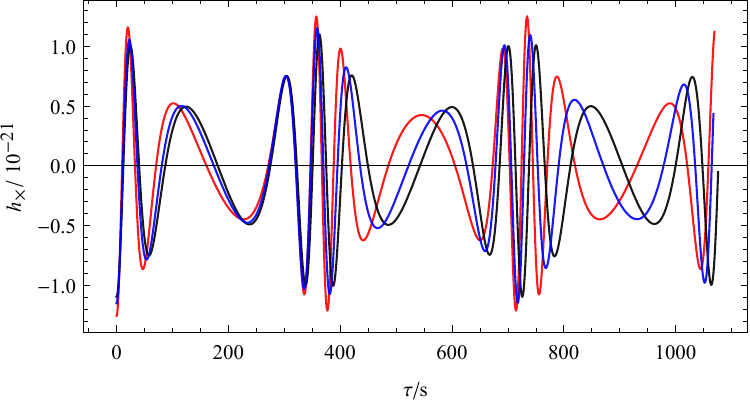}\caption{Right panel: Combined $h_+$ waveforms from Fig.~\ref{h+411}. Left panel: Combined $h_{\times}$ waveforms from Fig.~\ref{h*311}.}\label{h411combine}
\end{figure*}

In Figs.~\ref{h+311}-\ref{h411combine}, the periodic orbits characterized by the integer triplet $(z,w,v)$ exhibit distinct zoom and whirl phases within each complete cycle. During the zoom phase, the particle moves along an extended, elliptical segment of the orbit, travelling farther from the BH, where the gravitational field is comparatively weaker. This motion produces the quieter portions of the GW signal, visible in both the $h_+$ and $h_{\times}$ polarizations, and corresponds to the leaf-like structures of the orbit. As the particle spirals inward, it transitions into the whirl phase, during which it executes multiple tight, nearly circular loops in the strong-field region near the BH. This behavior generates sharp, high-amplitude bursts in the GW signal due to the stronger gravitational interaction. The number of quiet intervals in the waveform reflects the number of zoom leaves, while the number of loud bursts corresponds to the number of whirls. Thus, the waveform structure encodes the orbital characteristics specified by the integer triplet $(z,w,v)$.

Compared with the results presented in \cite{2025EPJC...85.1340Z}, we observe that the overall structure of the GW signals, characterized by alternating quiet zoom phases and sharp whirl-induced bursts, exhibits a similar qualitative behavior. However, in our analysis, the electric charge $Q$ in the ENLMY BH introduces an additional modification to the orbital dynamics. As a consequence, the resulting GW amplitudes and oscillation patterns exhibit slight deviations from those of the simple Yukawa BH \cite{2025EPJC...85.1340Z}.

\section{Fundamental frequencies \label{sec.freq}}

In this section, we investigate the characteristic frequencies of test particles orbiting the ENLMY BH described by the spherically symmetric line element Eq.(\ref{eq.ds}), where the metric function $f(r)$. In particular, we derive the Keplerian (orbital) frequency $(\Omega_k)$ of circular motion, as well as the radial $(\Omega_r)$ and vertical epicyclic $(\Omega_\theta)$ frequencies associated with small perturbations around stable circular orbits in the equatorial plane. These fundamental frequencies provide an important tool to probe the dynamical properties of the ENLMY spacetime. They can be directly applied to model the observed upper and lower QPO frequencies in accreting compact objects.

\subsection{Keplerian frequency}

The angular momentum of test particles in circular orbits (so-called) the Keplerian orbits $\Omega_k=\frac{d\phi}{dt}$ for the  static ENLMY BH defined as \cite{Rayimbaev2021Galax}: 
\begin{eqnarray}\label{eq.Kep}
\nonumber
\Omega_k^2= \frac{M}{r^3}
    - \frac{q^2}{12r^3}\left[B'(r) - \frac{2B(r)}{r}\right].
\end{eqnarray}
To estimate the value of the fundamental frequencies, we express them in units of Hertz (Hz) as follows:
\begin{equation}\label{toHz}
\nu = \frac{1}{2\pi}\frac{c^3}{GM} \Omega\ , [{\rm Hz}]\ .
\end{equation}
Note that for Eq.(\ref{toHz}), the values of the speed of light in vacuum and the gravitational constant are $c=3\times 10^8$ m/sec and $G=6.67\times 10^{-11}\rm m^3/(kg\times sec^2)$, respectively.

\begin{figure}
\centering \includegraphics[width=0.99\linewidth]{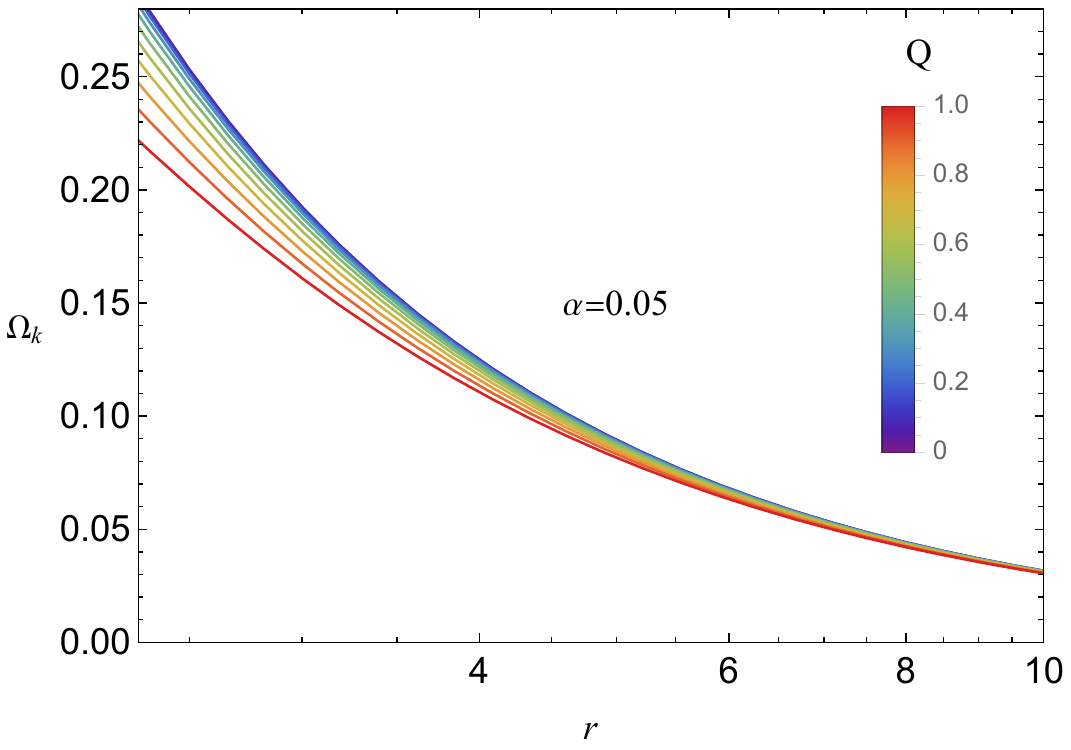}
\includegraphics[width=0.99\linewidth]{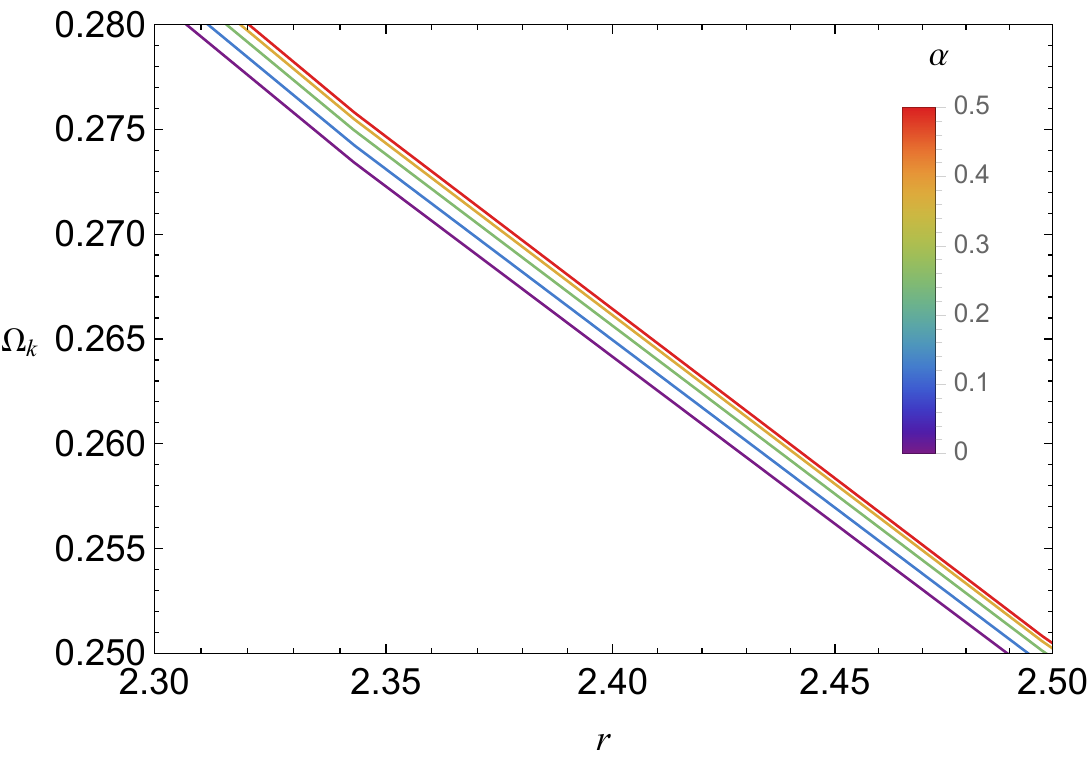}
\caption{Keplerian frequencies}
\label{fig.Kep} 
\end{figure}

It can be shown from top panel of  Fig.\ref{fig.Kep} that, the Yukawa contribution is fixed at $\alpha=0.05$, so the spacetime behaves close to a RN--type deformation.  
As $Q$ increases, the term $-\frac{Q^2}{6r^{2}} B(r)$ introduces an increasingly negative contribution to the lapse function, thereby reducing both $f(r)$ and its radial derivative.  Consequently, the Keplerian frequency decreases monotonically with increasing $Q$ within the given range. The suppression is strongest in the inner region of the accretion disc, where the $Q^{2}/r^{2}$ dependence amplifies the modification. At the same time, at large radii all curves converge, reflecting the rapid decay of the electromagnetic contribution at asymptotic distances. This behaviour reveals that the electric charge predominantly affects the strong-gravity regime near the BH.

The bottom panel of this graph presents the dependence of $\Omega_k$ on the radial coordinate for different values of the Yukawa screening parameter. Here, the exponential and integral components of $B(r)$ become dominant, and the Yukawa sector alters the slope of the lapse function in a qualitatively different manner than the charge. Increasing $\alpha$ enhances the effective Yukawa modification, leading to a systematic increase in $f'(r)$ within the given range. As a result, $\Omega_k$ rises monotonically with $\alpha$, and the nearly parallel arrangement of the curves indicates a smooth and regular dependence on the screening strength, indicating that Yukawa corrections strengthen the effective gravitational interaction.

\subsection{Harmonic ossilations}
Next, we consider small perturbations to test-particle motion around equatorial circular geodesics and introduce the radial and vertical epicyclic frequencies, $\Omega_r$ and $\Omega_\theta$, respectively. For the ENLMY metric Eq.(\ref{ENLMYmetric}), the radial epicyclic frequency can be expressed in terms of the Keplerian frequency $\Omega_k$ as \cite{Rayimbaev2021Galax}:
\begin{equation}
    \left(\frac{\Omega_r}{\Omega_k}\right)^2
    = 3 f(r) - 2 r f'(r) + \frac{r f(r) f''(r)}{f'(r)}.
\end{equation}

\begin{figure}
    \centering
    \includegraphics[width=0.99\linewidth]{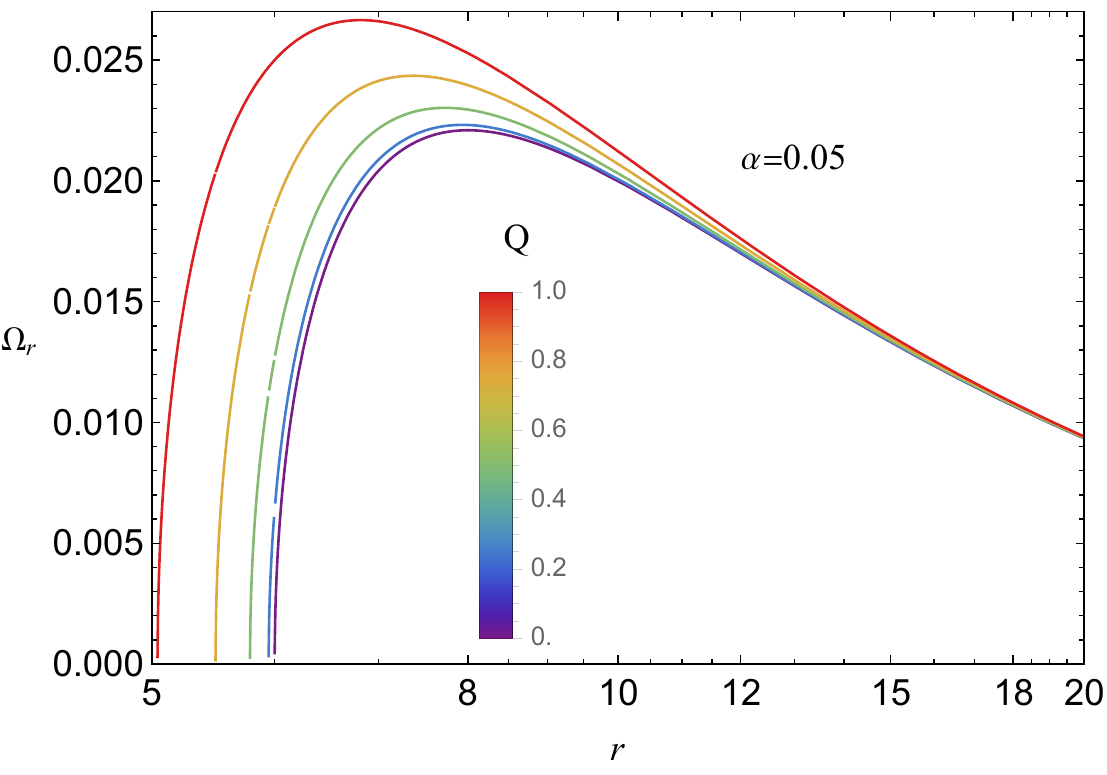}
    \includegraphics[width=0.99\linewidth]{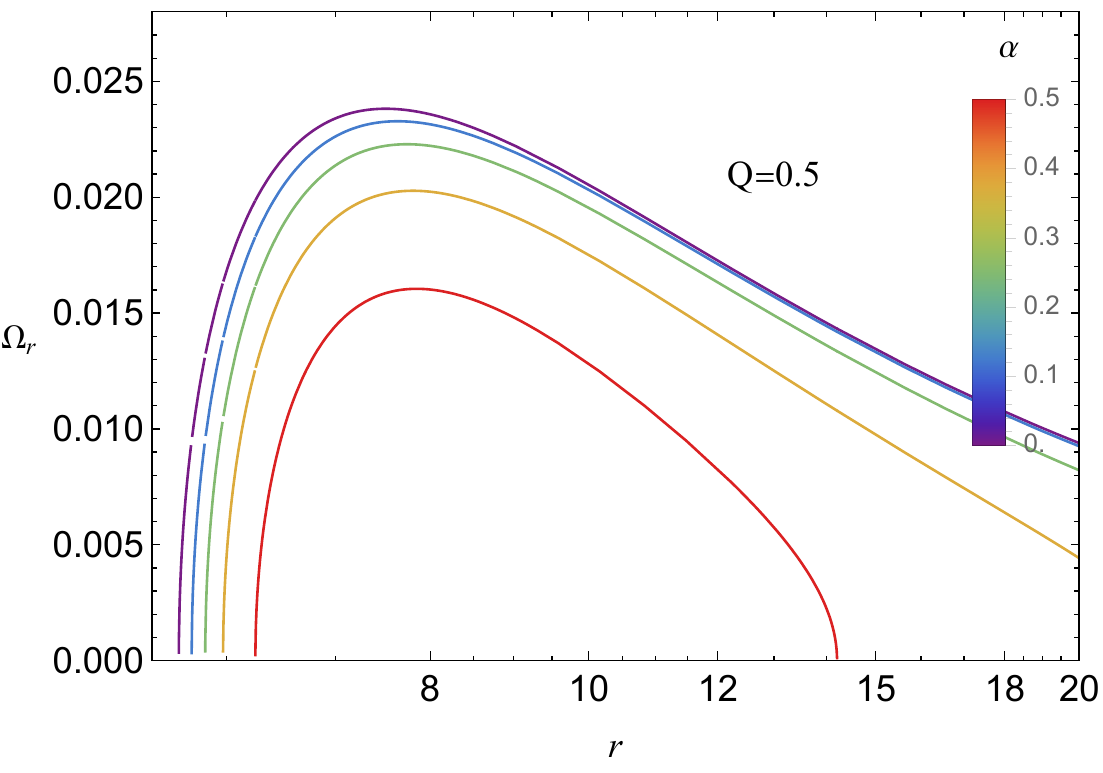}
    \caption{Radial epicyclic frequency as a function of the radial coordinate for several values of the $Q\ \& \ \alpha$ (top $\&$ bottom panels respectively) parameters.}
    \label{Fig.rad}
\end{figure}

Due to spherical symmetry, the vertical epicyclic frequency coincides with the Keplerian one
\begin{equation}
    \Omega_\theta = \Omega_k=\Omega_\phi.
\end{equation}

The top panel of Fig.\ref{Fig.rad} displays the radial epicyclic frequency for several values of the charge parameter while keeping $\alpha=0.05$.  The curves exhibit a well-defined maximum, whose position marks the transition region between the strongly curved inner geometry and the weak-field regime.  An increase in $Q$ leads to a uniform upward shift of the entire profile: the peak value of $\Omega_r$ becomes higher, and the maximum is displaced slightly inward, toward the BH. This behaviour indicates that the charge term in the ENLMY metric effectively modifies the radial curvature gradients. For larger values of $Q$, the combined contribution of $f(r)$ and its derivatives increases the effective potential and provides stronger radial oscillations at smaller radii. At sufficiently large distances, all curves merge, indicating that the influence of $Q$ rapidly diminishes outside the strong-gravity region.

The bottom panel of this figure shows the dependence of $\Omega_r$ and $r$ on the Yukawa parameter for different values of $ Q$ when the charge is held at $Q=0.5$. In contrast to the charge dependence, the Yukawa screening parameter induces a qualitatively different deformation of the radial oscillation frequency.  As $\alpha$ grows, the peak of $\Omega_r$ decreases and moves outward, away from the BH.  
This outward drift of the maximum reflects the increasing dominance of the exponential and integral terms in $B(r)$, which soften the curvature close to the object and push the region of strongest radial restoring force to larger radii.

For sufficiently large $\alpha$, the inner part of the profile is significantly suppressed, indicating the onset of weakened radial stability and a shift of the ISCO to larger radii.  
The progressive reduction of $\Omega_r$ at small $r$ therefore signals that Yukawa screening counteracts the radial restoring force supplied by the central potential.

Thus, increasing $\alpha$ lowers the maximal value of the epicyclic frequency and pushes its location outward into the weaker-field domain.

\section{Quasiperiodic oscillations }\label{QPOs}

QPOs represent some of the most informative timing signatures detected in electromagnetic radiation from compact objects, including BHs, neutron stars, white dwarfs, and their binary systems~\cite{Ingram2010MNRAS,Zdunik2000AA,Schaab1999MNRAS}. While the majority of confirmed QPOs arise in neutron-star binaries and other close binary systems, a smaller but still significant subset is observed in BH and white dwarf accretors. These oscillations are generally interpreted as (quasi-)harmonic motions of accreting matter in the innermost regions of the disk, where relativistic effects strongly influence particle dynamics. Because of this, QPOs act as sensitive tracers of orbital motion, epicyclic behavior, and the structure of the underlying spacetime. Their characteristic frequencies carry direct information about curvature, frame dragging, disk inhomogeneities, and potential deviations from general relativity. In particular, high-frequency twin-peak QPOs—typically modeled through combinations of orbital, radial, and vertical oscillation modes—provide a powerful tool for constraining key BH parameters such as mass, spin, and the role of additional surrounding fields~\cite{Torok2019MNRAS,Klis2000ARAA}.
In this section, twin-peak QPO frequencies around the BH are analysed by a comparison with the Schwarzschild case, RN$(Q=0.3)$  and a relationship between the upper $(\nu_U)$ and lower $(\nu_L)$ QPO frequencies is established within two theoretical frameworks: the RP model \cite{Stella1999ApJ} ( $\nu_U=\nu_\phi$ and $\nu_L=\nu_\phi-\nu_r$)   and the warped disc (WD) model $(\nu_U=2\nu_\phi-\nu_r,\  \nu_L=2\nu_\phi-2\nu_r)$ \cite{Kato2004PASJ,Kato2008PASJ}. 

\begin{figure*}[ht!]
\centering
\includegraphics[width=0.49\linewidth]{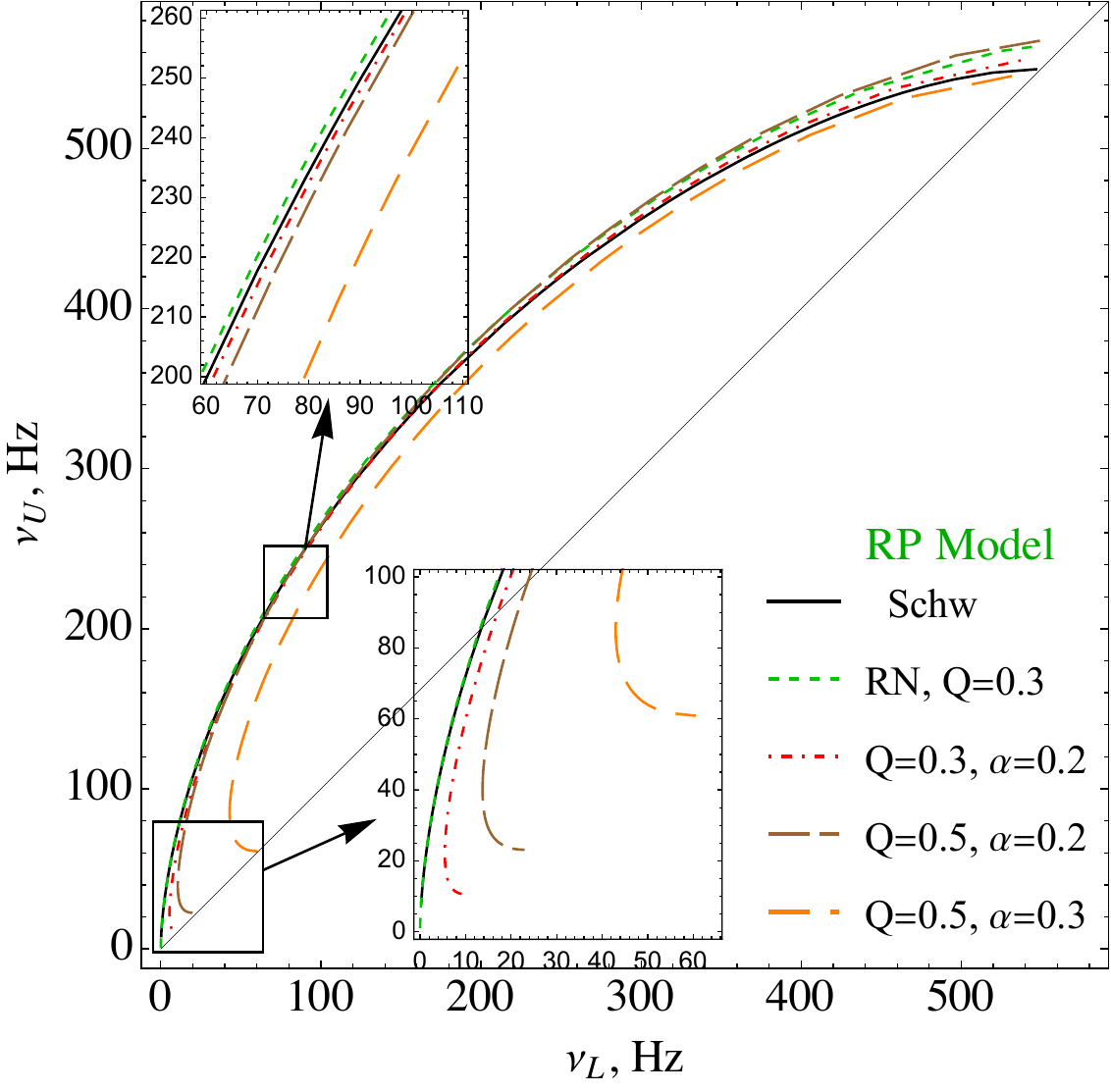}
\includegraphics[width=0.49\linewidth]{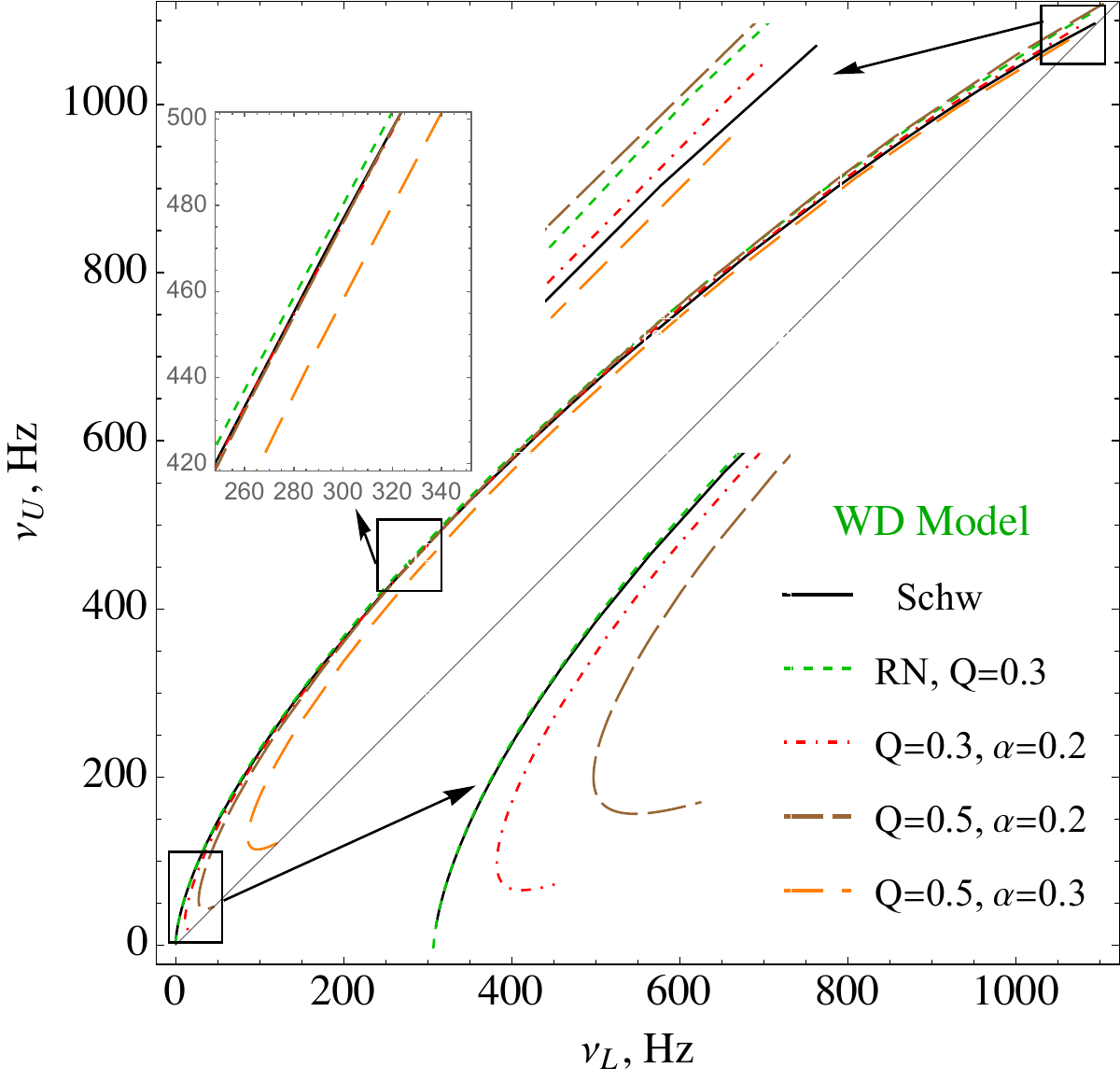}
\caption{Upper and lower frequencies observed around the BH}
\label{models}
\end{figure*}

Fig.\ref{QPOs} shows the upper and lower frequencies of a BH in RP and WD models corresponding to the change of the parameters $Q$, and $\alpha$. In summary, looking at each model, the following conclusions can be drawn:
\begin{itemize}
   \item In the low frequency range (long radial coordinates), the Schwarzschild and the RN cases overlap. For a non-zero Yukawa screening parameter, lines shift below the RN line. With $\alpha$ fixed, increasing $Q$ shifts, these lines even further down. This indicates that the motion of the test particle at large radial coordinates is affected by the Yukawa screening parameter, and the effect increases as the BH's charge rises.
   \item In the RP model, the RN curve lies above all the others in the mid-frequency region, while the remaining curves behave similarly to those in the low-frequency range. In the WD model, the RN curve is likewise positioned at the top, and for small values of $\alpha$ and $Q$ it remains almost identical to the Schwarzschild curve. For larger values of $Q$ and $\alpha$, however, the corresponding curve shifts to the lowest position in both models.
   \item In the high-frequency regime, the upper curve corresponds to configurations with larger values of the charge $Q$ and relatively small values of the Yukawa parameter $\alpha$. In the innermost regions of the accretion flow near the BH, the influence of electric charge becomes increasingly dominant over Yukawa screening. As a result, an increase in $Q$ enhances the orbital frequencies more efficiently than variations in $\alpha$ within this strong-gravity zone.
   \item Overall, in both models, while $\alpha$ tends to dominate at large radial coordinates, the Yukawa screening effect persists toward the BH, and the dominance of the electric charge $Q$ gradually increases as one approaches the strong-gravity region.
\end{itemize}

\subsection{Radius of QPO orbit \label{radiusQPO}}

In this subsection, we analyze the effect of the Yukawa screening parameter on the QPO radius, considering the ratios 3:2, 4:3, and 5:4. 
\begin{equation}
    \label{Eqrqpo}
 m\nu_L=n\nu_U\ ,
\end{equation}
where $n$ and $m$ are integers. %\cite{JRqpo2021,JRqpo2023uero,JR.2024qpoChPhy,JR.qpo2023EPJ,JR.mcmcqpo24,JR.galax.qpo,JR.qpo.gal2,JR.qpo.epjc23,JR.qpo.epjc23(2),JR.qpoAP.23, JR.qpo.epcj.22}.
One can show the behavior of the QPO orbits for the given resonant cases by contour-plotting Eq. (\ref{Eqrqpo}) for the fixed value $Q=0.3$.

\begin{figure*}[ht!]
\centering  
\includegraphics[width=0.49\linewidth]{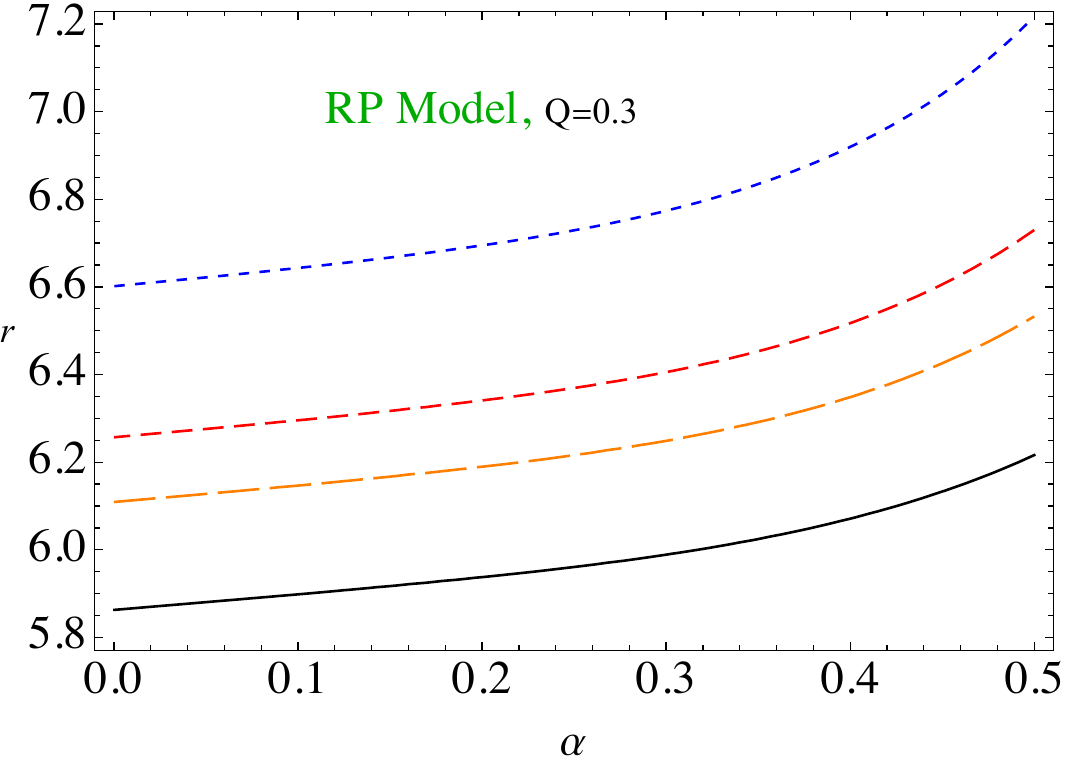}
\includegraphics[width=0.49\linewidth]{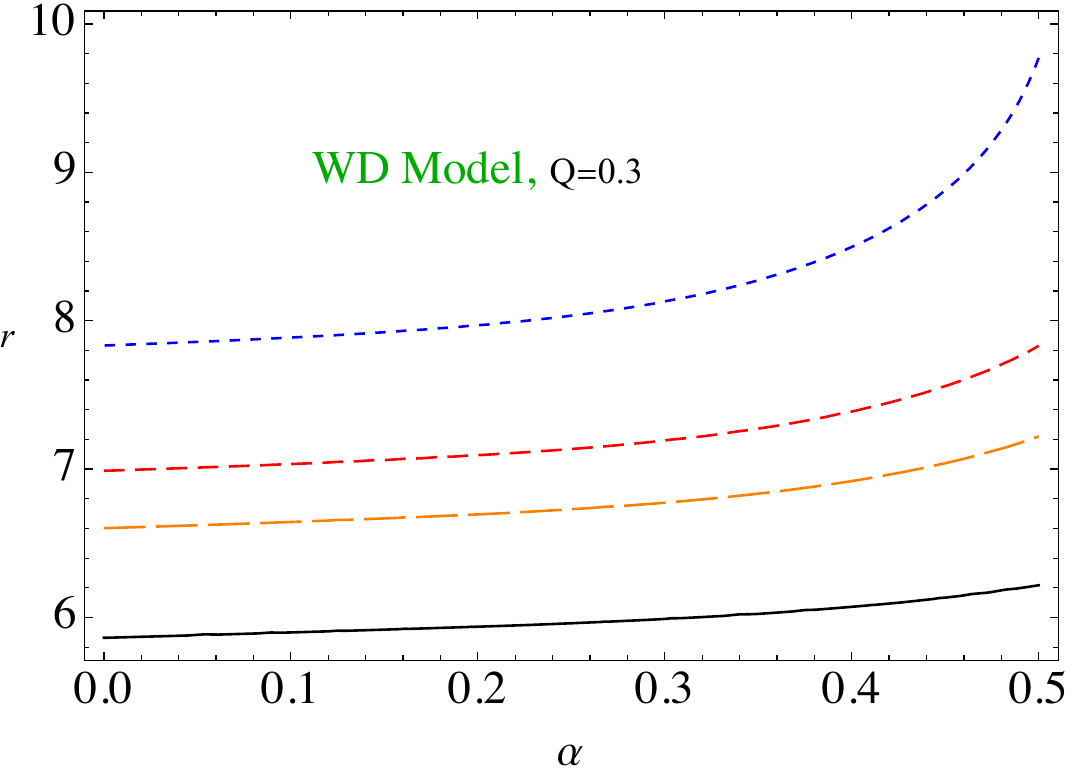}
\caption{$r$ depends on $\alpha$ which ISCO (black), $3:2$ (blue), $4:3$ (red), $5:4$ (orange) ratios.}
    \label{QPOs ratios}
\end{figure*}
From Fig.~\ref{QPOs ratios}, we can take various conclusions: 
\begin{itemize}
    \item As the $\alpha$ increases, the resonance QPO radius systematically shifts outward. This outward growth is more pronounced in the RP model, whereas in the WD model, it is noticeably slower.
    \item In comparison with the WD model, the resonance radii in the RP model lie much closer together, the interval $r_{RP}\in(5.8,\,7.2)$. In contrast, in the WD model they extend over a significantly broader range $r_{WD}\in(5.8,\,10)$.
    \item In both models, as the ratio $\nu_U\!:\!\nu_L$ approaches 1, the resonance radii approach the ISCO curve. A similar tendency was also observed in our earlier works \cite{2023AnPhy.45469335R,2022IJMPD..3140004R}.
\end{itemize}

\section{MCMC Analyses}\label{section6}

In this section, we explore a range of BH masses in order to constrain the Yukawa screening parameter $\alpha$ and the BH electric charge $Q$. We consider three representative BH classes, including stellar-mass and intermediate-mass BHs. The stellar-mass BH sample consists of the compact objects GRO~J1655--40, XTE~J1550--564, and GRS~1915+105. In addition, we use QPO observations from M82~X--1 \cite{pasham2014blackhole}, which is widely regarded as an intermediate-mass BH with an estimated mass in the range of $400$--$900\,M_\odot$, residing in X-ray source in the galaxy M82 \cite{2008MNRAS.387.1707C,2006ApJ...637L..21D,2006MNRAS.365.1123M,2010ApJ...712L.169F,2006MNRAS.370L...6P,2010ApJ...710...16Z}. The corresponding QPO frequencies for all considered sources are given in Table~\ref{table1}.

\begin{table*}[ht!]\centering
\begin{center}
\caption[]{\label{table1} QPOs frequencies observed in microquasars and Galactic centre}
%\begin{displaymath}
\renewcommand{\arraystretch}{1.2}
\begin{tabular}{| l || c  c | c  c | l |}
\hline
Source %\cite{Remillard2006,,,,} 
&  $\nu_{\rm{U}}\,$[Hz]&$\Delta\nu_{\mathrm{U}}\,$[Hz]& $\nu_{\rm {L}}\,$[Hz]&$\Delta\nu_{\rm{L}}\,$[Hz]& Mass [\,M$_{\odot}$\,] \\
\hline
\hline
XTE~J1550-564 \cite{2011ApJ...730...75O} & 276&$\pm\,3$& 184&$\pm\,5$&  ${9.1\pm0.61}$\\
GRO~J1655--40 \cite{Strohmayer2001}    & 451&$\pm\,5$& 298&$\pm\,4$ & {5.4$\pm$0.3}   \\
GRS~1915+105 \cite{2015ApJ...814...87M}   & 168&$\pm\,3$& 113&$\pm\,5$&  ${12.4^{+2.0}_{-1.8}}$   \\
M82 X-1 \cite{pasham2014blackhole}  & 5.07 & $\pm\,0.06$ & 3.32& $\pm\,0.06$ &$415\pm\,63$\\ 

\hline
            %\noalign{\smallskip}          \hline
        \end{tabular}
    %\end{displaymath}  
\end{center}
\end{table*}

To determine the best-fit values of the four model parameters using the QPO frequencies measured in microquasars, we employ a standard $\chi^{2}$ minimization procedure following the methodology outlined in \cite{2015EPJC...75..162B}:
\begin{eqnarray}\nonumber
\chi^{2}\left(M,\delta,\lambda,r\right)&=&\frac{(\nu_{1\phi}-\nu_{1\rm U})^{2}%
}{\sigma_{1\rm U}^2}+\frac{(\nu_{1\rm per}-\nu_{1\rm L})^{2}%
}{\sigma_{1 \rm L}^2}\\ \nonumber
&+&\frac{(\nu_{1\rm nod}-\nu_{1\rm C})^{2}%
}{\sigma_{1\rm C}^2}+\frac{(\nu_{2\phi}-\nu_{2\rm U})^{2}}%
{\sigma_{2\rm U}^2}\\
&+&\frac{(\nu_{2\rm nod}-\nu_{2\rm C})^{2}}%
{\sigma_{2\rm C}^2}~.
\end{eqnarray}

The optimal estimates of the parameters $M$, $\alpha$, $Q$, and the QPO orbital radius $r$ correspond to the configuration for which the $\chi^{2}$ function attains its minimum value, $\chi_{\rm min}^{2}$. The statistically preferred parameter ranges at a given confidence level (C.L.) are defined by the interval $\chi_{\rm min}^{2}\pm \Delta\chi^{2}$. By minimizing the $\chi^{2}$ function, we extract the mean (best-fit) values of $M$, $\alpha$, $Q$, and $r$, which are given in Table~\ref{prior}.

\begin{table*}[ht!]\centering
\begin{center}
\renewcommand\arraystretch{1.5} 
\caption{\label{prior}%
The Gaussian prior of BH spacetime QPOs for the microquasars and the galactic center.}
% \begin{ruledtabular}
\begin{tabular}{lcccccccccc}
\hline\hline
\multirow{2}{*}{$P$} & \multicolumn{2}{c}{XTE J1550-564} & \multicolumn{2}{c}{GRO J1655-40} & \multicolumn{2}{c}{GRS 1915+105} & \multicolumn{2}{c}{M82 X-1} \\
& $\mu$ & \multicolumn{1}{c}{$\sigma$} & $\mu$          & $\sigma$ & $\mu$ & \multicolumn{1}{c}{$\sigma$} & $\mu$ & \multicolumn{1}{c}{$\sigma$} \\
\hline
     $M/M_{\odot}$ & $6.691$ & 0.109 & $4.068$ & $0.093$ & $11.061$ & $0.109$ & $359.62$ & $73.15$ \\
    
     $\alpha\times10^2$ & 9.99 & 1.45   & $10.87$  & $1.53$ & 10.38 & 1.13 & $9.72$ & $1.07$  \\ 
    
     $Q$ & 0.092 & 0.023 & $0.098$ & $0.037$ & $0.098$ & $0.037$ & 0.095 & 0.017 \\
    
     $r$ & 6.766 & 0.129 & $6.737$ & $0.123$ & 6.737 & 0.123 & 6.795 & 0.132 \\
     \hline\hline
\end{tabular}
\end{center}
\end{table*}

We carried out a MCMC analysis using the \texttt{emcee} Python framework \cite{emcee} to constrain the dynamical parameters governing test-particle motion in the ENLMY BH spacetime. The modeling of quasi-periodic oscillations is performed within the relativistic precession (RP) model (see Section~\ref{QPOs}), following Refs.~\cite{Zahra:2025fvq,Liu-etal2023},

The corresponding posterior probability distribution is defined as \cite{Liu-etal2023},

\begin{eqnarray}
\mathcal{P}(\theta |\mathcal{D},\mathcal{M})=\frac{P(\mathcal{D}|\theta,\mathcal{M})\ \pi (\theta|\mathcal{M})}{P(\mathcal{D}|\mathcal{M})},
\end{eqnarray}
where $\pi(\theta)$ is the prior and $P(D|\theta,M)$ is the likelihood. We adopt Gaussian (normal) prior distributions for all model parameters, restricted to physically motivated bounds as summarized in Table~\ref{prior}. Specifically, the priors are defined as
$\pi(\theta_i) \sim \exp\!\left[-\frac{1}{2}\left(\frac{\theta_i-\theta_{0,i}}{\sigma_i}\right)^2\right], \qquad \theta_{\text{low},i}<\theta_i<\theta_{\text{high},i},$
where $\theta_{0,i}$ and $\sigma_i$ denote the mean values and standard deviations of the corresponding parameters. In the present analysis, the parameter vector is given by $\theta_i=\{M,\alpha,Q,r\}$. Guided by the expressions for the orbital and periastron precession frequencies discussed in Section~\ref{QPOs}, we employ distinct observational data subsets within the MCMC framework. Under these assumptions, the resulting likelihood function $\Lambda$ can be written as:
\begin{eqnarray}
\log \Lambda = \log \Lambda_{\rm U} + \log \Lambda_{\rm L},\label{likelyhood}
\end{eqnarray}
where $\log \Lambda_{\rm U}$ denotes the likelihood of the astrometric upper/orbital frequencies,
\begin{eqnarray}
 \log \Lambda_{\rm U} = - \frac{1}{2} \sum_{i} \frac{\left(\nu_{\phi\rm, obs}^i -\nu_{\phi\rm, th}^i\right)^2}{\left(\sigma^i_{\phi,{\rm obs}}\right)^2} \ ,
\end{eqnarray}
and $\log \Lambda_{\rm L}$ represents the probability (likelihood) of the data of the lower or periastron precession frequency ($\nu_{\rm per}$).

\begin{eqnarray}
\log \Lambda_{\rm L} =-\frac{1}{2} \sum_{i} \frac{\left(\nu_{\rm per, obs}^i -\nu_{\rm per, th}^i\right)^2}{\left(\sigma^i_{\rm per,{\rm obs}}\right)^2}.\ 
%- \frac{1}{2} \sum_{i} \frac{(V_{\rm R, obs}^i - V_{\rm R, the}^i)^2}{(\sigma^i_{V_{\rm R, obs}})^2},
\end{eqnarray}

Here, $\nu^{i}_{\phi,\mathrm{obs}}$ and $\nu^{i}_{\mathrm{per},\mathrm{obs}}$ denote the observed values of the orbital (Keplerian) frequency $\nu_{\mathrm{K}}$ and the periastron precession frequency, defined as $\nu_{\mathrm{per}}=\nu_{\mathrm{K}}-\nu_{r}$, for the astrophysical sources under consideration. Correspondingly, $\nu^{i}_{\phi,\mathrm{th}}$ and $\nu^{i}_{\mathrm{per},\mathrm{th}}$ represent the theoretical predictions of these frequencies derived within the adopted spacetime model.

We subsequently carry out an MCMC exploration to constrain the parameter set $(M,\alpha,Q,r)$ characterizing the ENLMY black hole spacetime. The analysis adopts Gaussian prior distributions, motivated by parameter estimates reported in previous QPO studies. For each parameter, we draw approximately $5\times10^{4}$ samples from the corresponding prior distribution, allowing for an efficient exploration of the admissible region of the parameter space within the prescribed bounds and facilitating the identification of the statistically preferred parameter values.

\begin{figure*} \centering
\includegraphics[width=0.45\linewidth]{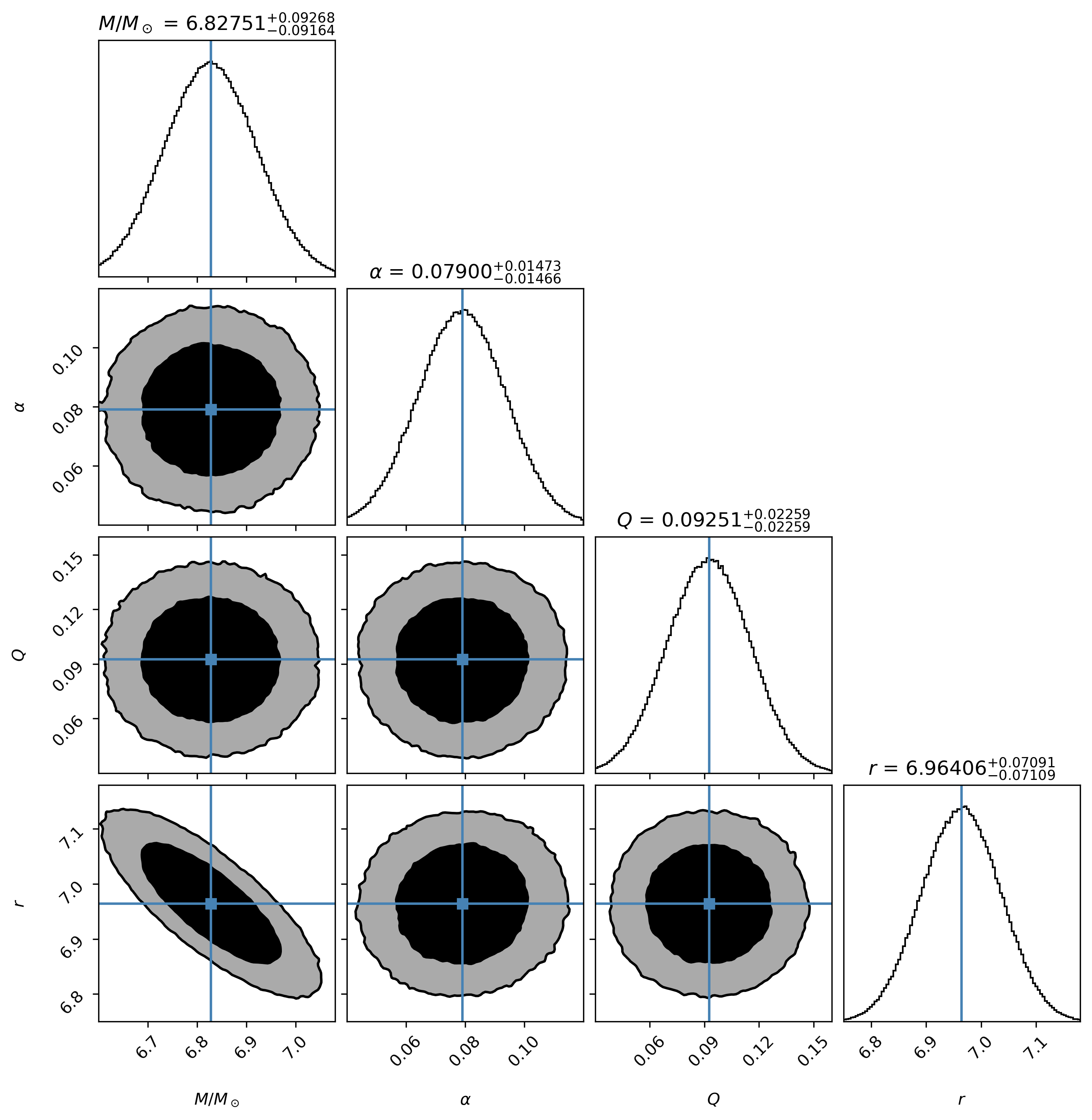}
\includegraphics[width=0.45\linewidth]{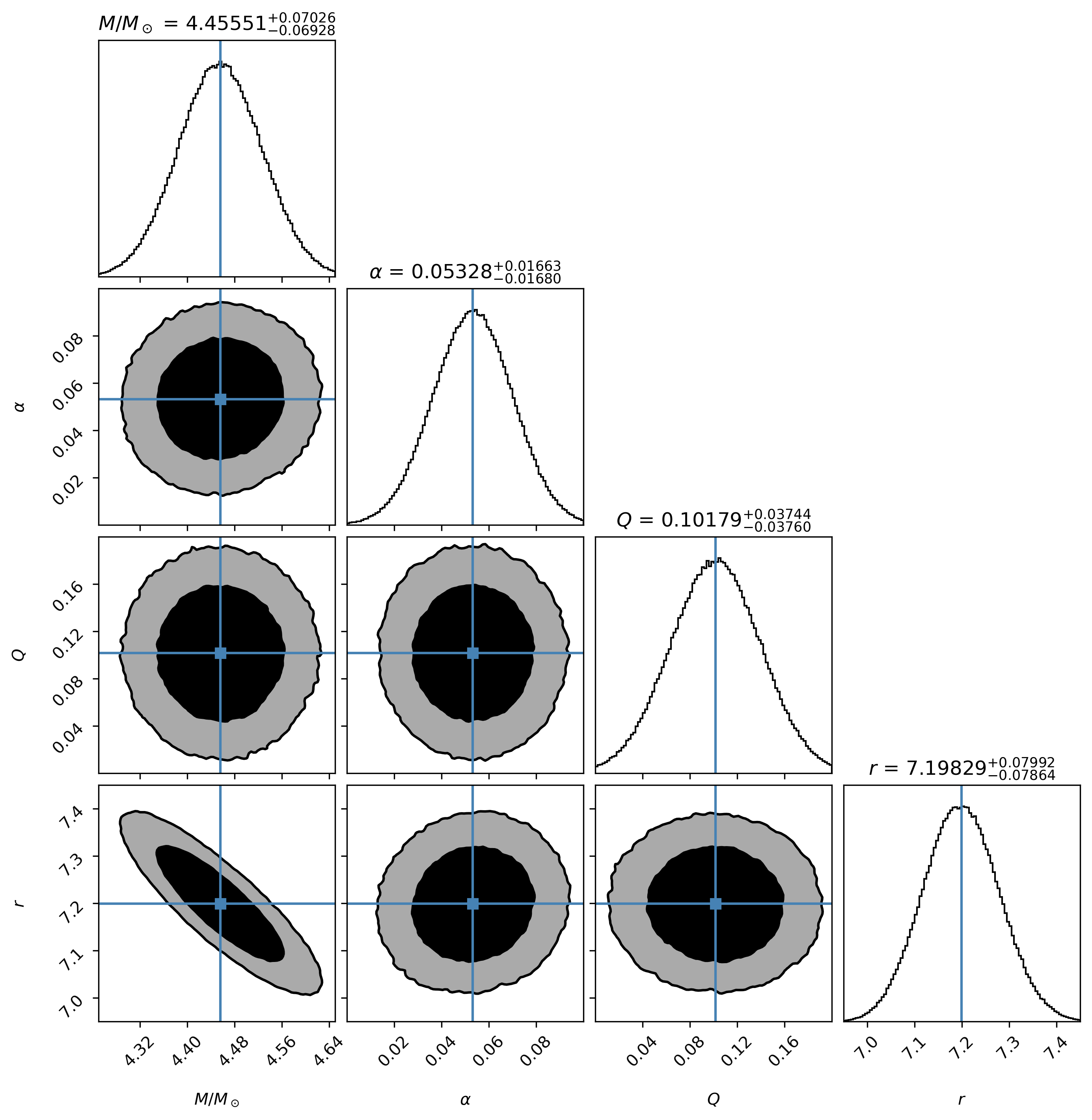}
\includegraphics[width=0.45\linewidth]{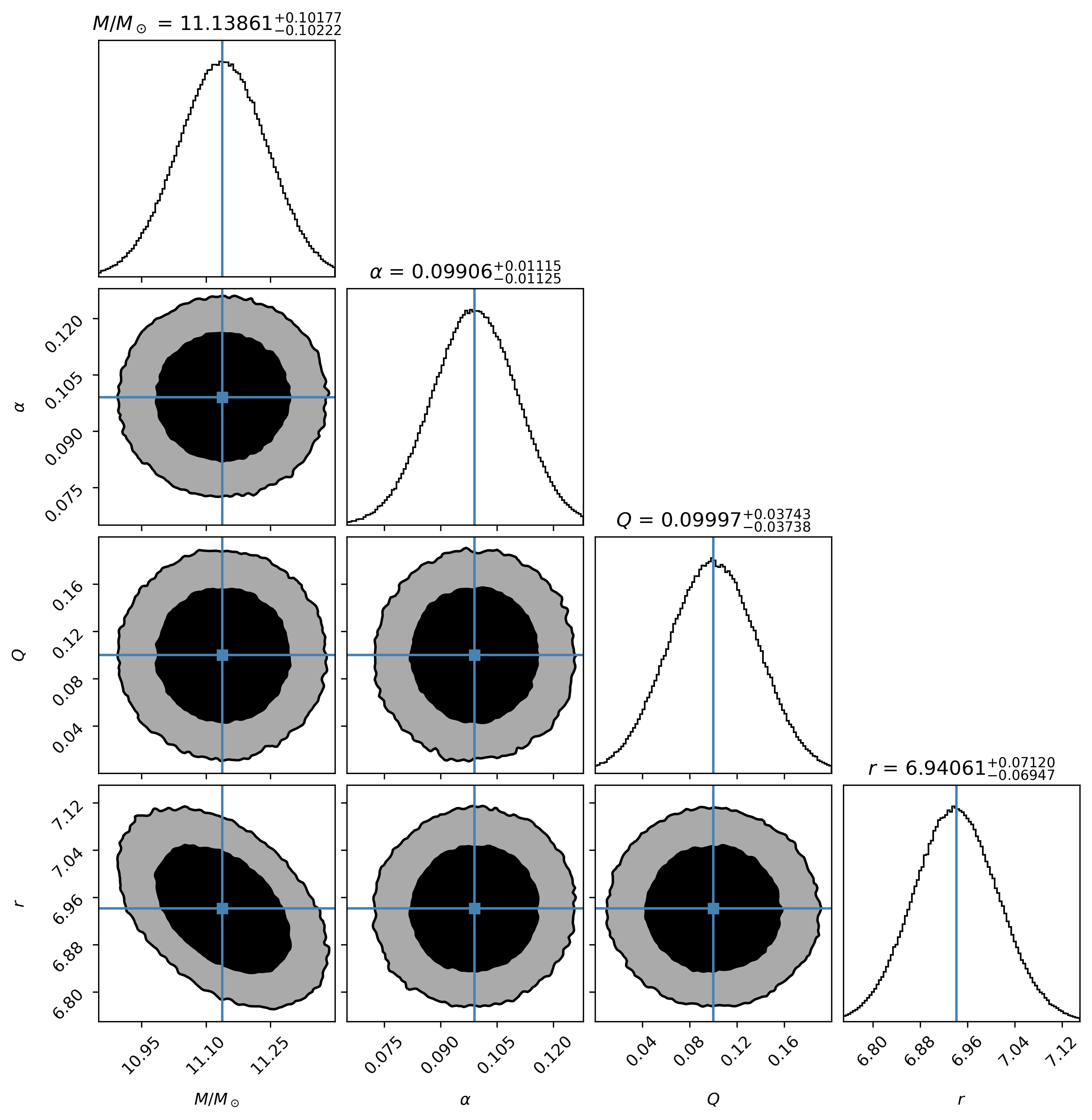}
\includegraphics[width=0.45\linewidth]{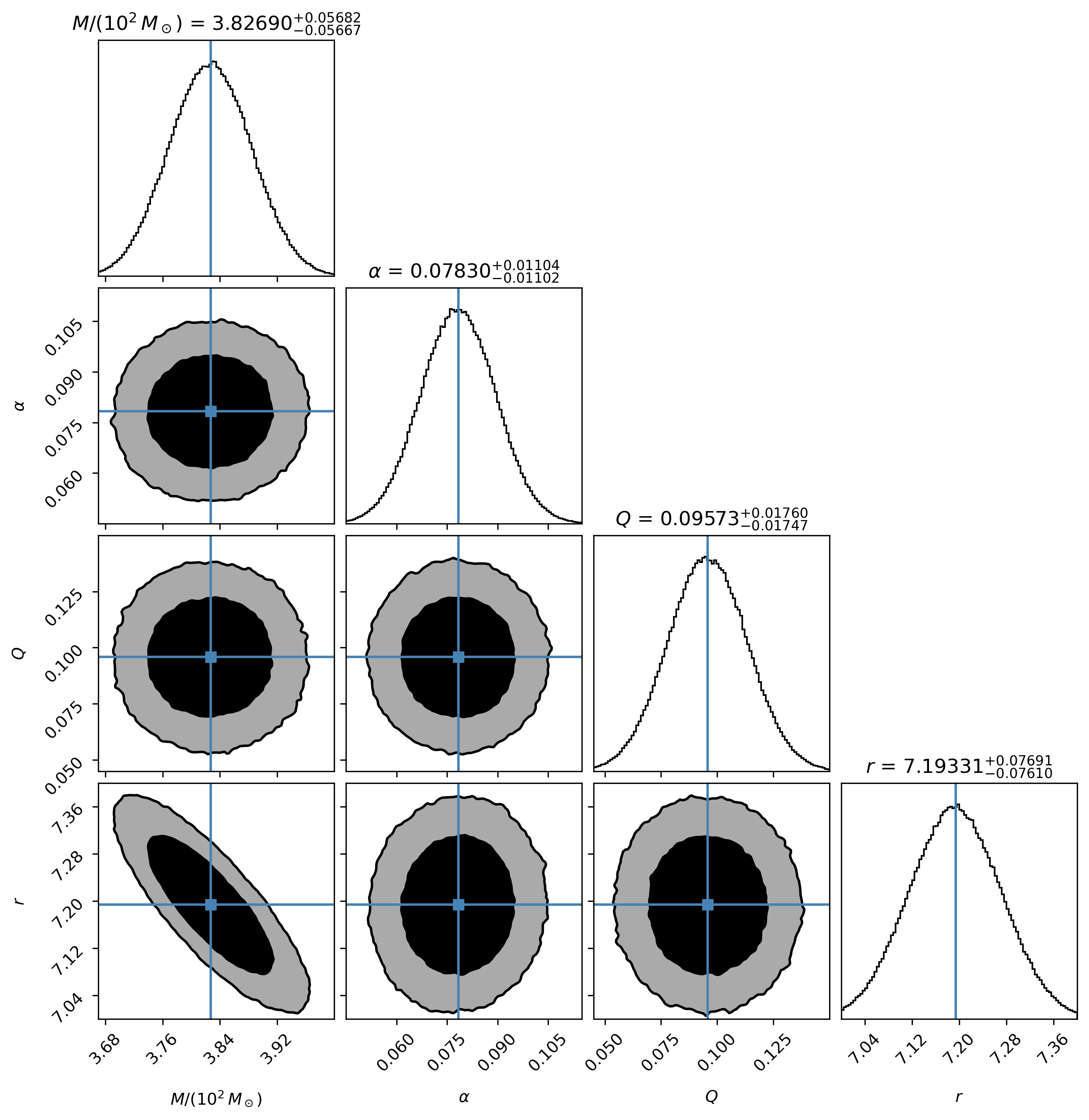}
\caption{Constraints on the BH mass, the Yukawa parameter, the BH charge and the radius of the QPO orbit from a four-dimensional MCMC analysis using the QPO data for the stellar-mass BH XTE J1550-564 (top left), GRO J1655-40 (top right), GRS 1915+105 (bottom left), and intermediate mass BH M82 X--1 (bottom right) in the RP model. \label{XTESgrA}} \end{figure*}

We summarize the best-fitting estimates of the parameters $(M,\alpha,Q,r)$ inferred from the MCMC posterior distributions, illustrated in Fig.~\ref{XTESgrA}, in tabular form.

\begin{table*} \centering \renewcommand\arraystretch{1.5} \caption{The best-fit values of the BH in ENLMY spacetime.} \label{tab:best_fit_value} 
\begin{tabular}{lccccc} \hline\hline \multicolumn{1}{c}{$P$} & \multicolumn{1}{c}{XTE J1550-564} & \multicolumn{1}{c}{GRO J1655-40} & \multicolumn{1}{c}{GRS 1915+105} & \multicolumn{1}{c}{M82 X-1} \\ \hline 
$M/M_{\odot}$ & $6.827^{+0.092}_{-0.091}$ & $4.455^{+0.070}_{-0.069}$ & $11.138^{+0.101}_{-0.102}$ & $3.826^{+0.056}_{-0.056} \times 10^2$ \\ 

$\alpha$ & $0.079^{+0.014}_{-0.014}$ & $0.053^{+0.016}_{-0.017}$ & $0.099^{+0.011}_{-0.011}$ & $0.078^{+0.011}_{-0.011}$  \\ 

$Q$ & $0.092^{+0.022}_{-0.022}$ & $0.102^{+0.037}_{-0.037}$ & $0.099^{+0.037}_{-0.037}$ & $0.095^{+0.017}_{-0.017}$\\ 

$r/M$ & $6.964^{+0.071}_{-0.071}$ & $7.198^{+0.079}_{-0.078}$ &  $6.941^{+0.071}_{-0.069}$ & $7.194^{+0.077}_{-0.076}$  \\ \hline\hline \end{tabular} \end{table*}

 The inferred BH masses show only marginal deviations when compared with the bounds obtained in our previous analysis and those reported in Refs.~\cite{2025EPJC...85..126J,2024EPJC...84..964J,2024ChJPh..92..143R,2024EPJC...84.1114R,2023Univ....9..391M,2023Galax..11..113R,2024ChPhC..48e5104R,2024JHEAp..44...99C,2024PDU....4601708M,2024PDU....4601561M}. This behavior can be attributed to the fact that QPO frequencies are governed not only by the BH mass but also by additional gravitational parameters. Consequently, constraints derived from QPO modeling naturally exhibit model-dependent variations.

It is worth emphasizing that, for nearly all parameters, the adopted prior distributions (Table~\ref{prior}) and the corresponding posterior estimates (Table~\ref{tab:best_fit_value}) are found to be in close agreement. This consistency indicates a weak prior dependence of the inference and suggests that the RP model provides a robust and reliable framework for describing QPOs in both microquasars and intermediate-mass BHs considered in this study.

%Our estimated masses exhibit minor deviations relative to the limits reported in our earlier study as well as those presented \cite{2025EPJC...85..126J,2024EPJC...84..964J,2024ChJPh..92..143R,2024EPJC...84.1114R,2023Univ....9..391M,2023Galax..11..113R,2024ChPhC..48e5104R,2024JHEAp..44...99C,2024PDU....4601708M,2024PDU....4601561M}. The oscillation frequencies depend on not only the BH mass but also gravity parameters, in particular, the Yukawa parameter $\alpha$. Therefore, the results of the QPO frequency constraint should also differ across cases.

\section{Conclusion}
In this work, we have investigated the bound orbits, GW emission, and related astrophysical signatures of test particles around the ENLMY BH. By analyzing the effective potential, we identified the radii of the ISCO and the IBCO, and examined how the Yukawa screening parameter $\alpha$ and electric charge $Q$ influence orbital stability. Our results indicate that increasing $\alpha$ weakens gravitational attraction at short distances, thereby increasing the angular-momentum requirement for circular motion and reducing the range of radii that can support stable orbits. Similarly, increasing the BH charge $Q$ introduces a repulsive effect, shifting both the ISCO and IBCO outward and lowering the particle energy required for circular orbits. These findings demonstrate that the ENLMY corrections significantly modify the classical orbital structure compared to the Schwarzschild spacetime.

We further explored the structure of periodic orbits between the ISCO and IBCO by classifying them using the integer triplet $(z,w,v)$. Our analysis shows that the orbits exhibit distinct zoom and whirl phases. The energy values associated with these periodic orbits are generally lower than those required in the Schwarzschild spacetime, particularly for moderate values of $\alpha$ and $Q$. However, for higher values of $\alpha$, the region between the ISCO and IBCO becomes too constrained to support closed periodic orbits, indicating that strong Yukawa screening limits the formation of stable periodic trajectories. 

Finally, we computed the GW radiation emitted by these periodic orbits using the quadrupole approximation. The resulting $h_+$ and $h_{\times}$
 waveforms clearly capture the zoom–whirl dynamics: quiet intervals correspond to the particle’s motion along extended, weak-field portions of the orbit, while high-amplitude bursts arise during tight whirls near the BH. These waveforms encode the orbital structure, providing a direct mapping from the integer triplets $(z,w,v)$ to observable GW signals. 

In comparison with \cite{2025EPJC...85.1340Z}, our results confirm that the qualitative features of periodic orbits and their associated GW signatures, particularly the characteristic zoom–whirl structure, remain consistent across different modified gravity frameworks. Nonetheless, including the electric charge $Q$ in the ENLMY BH yields noticeable quantitative differences.

Based on analyses of oscillation frequencies and QPO resonances, we identified two distinct gravitational regions. In the outer parts of the accretion disk, the Yukawa screening parameter \(\alpha\) has the strongest effect, lowering orbital frequencies. Closer to the BH, the electric charge \(Q\) becomes more important, increasing the orbital frequencies the most. This indicates that while the long-range Yukawa screening shapes the dynamics at large distances, near-horizon physics is increasingly governed by the BH's charge. The behavior of resonance radii for QPOs further highlights the distinct role of the Yukawa parameter in each model. As \(\alpha\) increases, the resonance radius shifts outward in both models, though this shift is more pronounced in the RP model than in the WD model. Notably, resonance radii in the RP model are confined to a relatively narrow range \(r_{\text{RP}} \in (5.8,\,7.2)\), whereas in the WD model they span a much broader interval \(r_{\text{WD}} \in (5.8,\,10)\). In both cases, as the ratio of the upper to lower QPO frequencies \(\nu_U\!:\!\nu_L\) approaches unity, the resonance radii converge toward the ISCO.

The MCMC analysis to estimate BH parameters from QPO data yields masses that exhibit slight, but systematic deviations from those reported in the literature. In particular, the results are slightly smaller than the observational data for the estimated masses of the quasars XTE J1550-564, GRO J1655--40, GRS 1915+105, and slightly larger for M82 X--1. The radial coordinate of particles for almost all BHs is around $r\sim7M$. This directly confirms that the QPO frequencies are not solely determined by mass but depend critically on the modified-gravity parameters. Therefore, any attempt to infer fundamental BH properties from QPO observations is inherently model-dependent. This work underscores the need to test and incorporate specific modified gravity frameworks into the analysis of high-energy astrophysical data.

\bibliographystyle{apsrev4-1}
\bibliography{main}

\end{document}